\documentclass[aps,prl,twocolumn,showpacs,preprintnumbers]{revtex4}
\usepackage{amsmath}
\usepackage{dcolumn}
\usepackage{bm}
\usepackage{subfigure}
\usepackage{amsfonts}
\usepackage{amssymb}
\usepackage{makeidx}
\usepackage{epsfig}
\usepackage{graphicx}

\begin{document}

\title{\textbf{Theory of the 1-point PDF for incompressible Navier-Stokes
fluids}}
\author{M. Tessarotto\thanks{%
Electronic-mail: M.Tessarotto@cmfd.univ.trieste.it}$^{a,b}$ and C. Asci$^{a}$%
}
\affiliation{$^{a}$Department of Mathematics and Informatics, University of Trieste, Italy%
\\
$^{b}$Consortium of Magneto-fluid-dynamics, University of Trieste, Italy}
\date{\today }

\begin{abstract}
Fundamental aspects of fluid dynamics are related to construction
of statistical models for incompressible Navier-Stokes fluids. The
latter can be considered either \textit{deterministic} or
\textit{stochastic,} respectively for \textit{regular} or
\textit{turbulent flows.} In this work we claim that a possible
statistical formulation of this type can be achieved by means of
the 1-point (local) velocity-space probability density function
(PDF, $f_{1}$) to be determined in the framework of the so-called
inverse kinetic theory (IKT). There are several important
consequences of the theory. These include, in particular, the
characterization of the initial PDF [for the statistical model
$\left\{ f_{1},\Gamma \right\} ]$ . This is found to be generally\
non-Gaussian PDF, even in the case of flows which are regular at
the initial time. Moreover, both for regular and turbulent flows,
its time evolution is provided by a Liouville equation, while the
corresponding Liouville operator is found to depend only on a
finite number of velocity moments of the same PDF. Hence, its time
evolution depends (functionally) solely on the same PDF. In
addition, the statistical model here developed determines uniquely
both the initial condition and the time evolution of $f_{1}.$ As a
basic implication, the theory allows the \textit{exact
construction of the corresponding statistical equation for the
stochastic-averaged PDF }and the \textit{unique representation of
the multi-point PDF}'s solely in terms of the 1-point PDF. As an
example, the case of the reduced 2-point PDF's, usually adopted
for the statistical description of NS turbulence, is considered.
\end{abstract}

\maketitle

\bigskip


\section{1 - Introduction}

The description of fluids, and more generally of continua, is based on the
introduction of a suitable set of \textit{fluid fields} $\left\{ Z\right\} $%
\textit{\ which define the state of each fluid }and obey, by assumption, a
well-posed set of PDE's denoted as \textit{fluid equations. }The formulation
appropriate for an incompressible Navier-Stokes (NS) fluid - based on the
so-called incompressible NS equations (INSE) - is recalled in Appendix A,
together with the basic notations and definitions here adopted. \ The fluid
equations, the fluid fields\ and the related initial and boundary
conditions, are considered either as \textit{deterministic} or \textit{%
stochastic, }respectively for\textit{\ regular }or \textit{turbulent }flows
(see Appendix A, Subsections A.2 and A.3).

The statistical treatment of fluids usually adopted for turbulent flows
(which may be invoked, however, to describe also regular flows) consists,
instead, in the introduction of appropriate axiomatic approaches denoted as
\textit{statistical models.} These are sets $\left\{ f,\Gamma \right\} $
formed by a suitable probability density function (PDF) and a phase-space $%
\Gamma $ (subset of $%
\mathbb{R}
^{n}$) on which $f$ is defined.\ By definition, a statistical model $\left\{
f,\Gamma \right\} $ of this type must permit the representation, via a
suitable mapping%
\begin{equation}
\left\{ f,\Gamma \right\} \Rightarrow \left\{ Z\right\} ,
\label{relationship}
\end{equation}%
of \textit{the complete set }(or more generally only\textit{\ of a subset})%
\textit{\ }of the \textit{fluid fields} $\left\{ Z\right\} \equiv \left\{
Z_{i},i=1,n\right\} $ which define the state of the same fluids. Their
construction involves, besides the specification of the phase space ($\Gamma
$) and the \textit{probability density function }(PDF) $f,$ the
identification of the functional class to which $f$\textit{\ }must belong,
denoted as $\left\{ f\right\} .$ As a consequence, the fluid fields $%
Z_{i}\in \left\{ Z\right\} $ are expressed in terms of suitable functionals
(called \textit{moments}) of $f.$ In the case of an incompressible (and
isentropic) Navier-Stokes fluid (INSF) the latter are identified with $%
\left\{ Z\right\} \equiv \left\{ \mathbf{V},p,S_{T}\right\} ,$ $\mathbf{V}$
and $p$ indicating respectively the fluid velocity, the fluid pressure, both
assumed strong solutions the incompressible Navier-Stokes equations (INSE),
and $S_{T}$ the constant thermodynamic entropy. As an alternative, $p$ can
be replaced by the kinetic pressure $p_{1},$ to be suitably defined [see
subsequent Eq.(\ref{kinetic pressure})].

Goal of this paper is to propose a \textit{statistical model} for
incompressible NS fluids, described by INSE, which is simultaneously \textit{%
(a) unique, (b) statistically complete }and\textit{\ (c) closed.}\ While the
precise meaning of these statements, in particular the relationship (\ref%
{relationship}), will be discussed in detail below, we anticipate that:

\begin{itemize}
\item the first condition (uniqueness) imposes that the same PDF and its
time evolution should both be uniquely prescribed in terms of \textit{%
conditional observables} (see related definitions in the Appendix A);

\item the second one (statistical completeness; see subsequent Subsection
1D) involves the assumption that the PDF should determine uniquely also a
prescribed set of \textit{physical observable} (or rather \textit{%
conditional observables}), to be suitably defined. The latter include in
particular the complete set of fluid fields describing the state of the
fluid;

\item the third one (closure) includes both \textit{moment- }and \textit{%
kinetic-closure conditions }(the corresponding definitions are given again
in Subsection 1D), involving \ - respectively - the requirements that there
exists a closed set of fluid equations and that the statistical equation
advancing in time the PDF depends\textit{\ }only the same PDF, via a finite
number of moments\textit{\ }(of the PDF)\textit{.}
\end{itemize}

\subsection{1A - CSM-inspired models for the $1-$point Liouville equation}

A well-known example of statistical model for incompressible NS fluids is
provided by the so-called \textit{statistical hydromechanics} developed
originally by Hopf \cite{Hopf1952} and later extended by Rosen \cite%
{Rosen1960} and Edwards \cite{Edwards1964} (\textit{HRE approach}). This
relies on the introduction of the $1-$\textit{point (or local)
velocity-space PDF, }$f_{1},$ to be intended as the conditional PDF of the
velocity $\mathbf{v}$ (\textit{kinetic velocity}) with respect the remaining
variables. In the HRE approach these are identified with $(\mathbf{r,}t),$
where $(\mathbf{r,}t)\in $ $\Omega \times I,$ while $f_{1}\equiv f_{1}(%
\mathbf{r,u,}t;Z),$ with $\mathbf{u\equiv v-V}(\mathbf{r,}t)$ the relative
kinetic velocity$,$ is identified with
\begin{equation}
f_{H}\equiv \delta \left( \mathbf{v-V}(\mathbf{r,}t)\right)
\label{DIRAC DELTA- HRE}
\end{equation}%
(\textit{deterministic PDF}), $f_{H}$ denoting the three-dimensional Dirac
delta defined in the velocity space $U\equiv
\mathbb{R}
^{3},$ with\ $\mathbf{v}$ belonging to $U\equiv
\mathbb{R}
^{3}$ and $(\mathbf{r,}t)\in \overline{\Omega }\times I$ (with $\overline{%
\Omega }$ denoting the closure of the configuration domain $\Omega \subseteq
\mathbb{R}
^{3}$). Hence, $f_{H}$ is defined by assumption on the set spanned by the
state vector $\mathbf{x=}(\mathbf{\mathbf{r,v}})\mathbf{,}$ i.e., $\overline{%
\Gamma }\equiv \overline{\Omega }\times U,$ with $\overline{\Gamma }$
denoting the closure of the \textit{restricted phase space}%
\begin{equation}
\Gamma \equiv \Omega \times U.  \label{restricted phase space}
\end{equation}%
It follows that, in this case, only the first two (velocity) moments of $%
f_{1},$ corresponding to $G=1,\mathbf{v,}$ are actually prescribed in terms
of the fluid fields and read
\begin{equation}
\int\limits_{U}d^{3}\mathbf{v}Gf_{H}(\mathbf{r,u,}t;Z)=1,\mathbf{V}(\mathbf{%
r,}t).  \label{moments of f_1 -HRE APPROACH}
\end{equation}%
Goal of the HRE approach in the case of turbulent flows is actually to
predict
\begin{equation}
\left\langle f_{1}(\mathbf{r,u,}t;Z)\right\rangle \equiv \left\langle f_{H}(%
\mathbf{r,u,}t;Z)\right\rangle ,  \label{f_1-HRE APPROACH}
\end{equation}%
the brackets $"\left\langle \cdot \right\rangle "$ denoting an ensemble
average, to be suitably prescribed, over the possible realizations of the
fluid \cite{Pope2000}. Its definition, however, is not unique. For example,
in the case of the so-called stationary and isotropic turbulence \cite%
{Batchelor} this is usually defined so that - by assumption - \ it commutes
with all the differential and integral operators (respectively, $\frac{%
\partial }{\partial t},\nabla ,$ $\nabla ^{2}$ and $\int\limits_{\Omega
}d^{3}r,$ $\int\limits_{U}d^{3}v$ and $\int\limits_{\Gamma }d^{6}x$)
appearing in the NS operator [see Eq.(\ref{NS operator}), in Appendix A]. As
an alternative, as discussed elsewhere \cite{Tessarotto2008-1},\ $%
\left\langle \cdot \right\rangle $ can also be identified with the
stochastic-averaging operator (\ref{stochastic averaging operator}) defined
in Appendix A.

Nevertheless, in the HRE approach $\left\langle f_{H}\right\rangle $ is not
directly determined. Rather, it is replaced by a suitable functional of $%
\left\langle f_{H}\right\rangle ,$ $\phi ,$ which obeys a
suitably-prescribed functional-differential equation (the so-called "$\phi $
equation" \cite{Hopf1952}). The problem of the construction of an equivalent
evolution (or so-called "transport") equation for $\left\langle
f_{H}\right\rangle $ has been investigated by several authors (see, for
example, Dopazo \cite{Dopazo}\ and Pope \cite{Pope2000}). The construction
of its formal solution is due to Monin \cite{Monin1967} and Lundgren \cite%
{Lundgren1967} (\textit{ML approach;} see also\ Monin and Yaglom, 1975 \cite%
{Monin1975} and therein cited references). However, alternative
(approximate) statistical approaches are known, such as the GLM (generalized
Langevin model; due to Pope \cite{Popo1983}). All of them typically rely on
the assumption of the existence of a suitable underlying \textit{phase-space}
\textit{classical dynamical system }which evolves in time the state vector $%
\mathbf{x}=(\mathbf{r,v}),$ namely the flow
\begin{equation}
T_{t_{o},t}:\mathbf{x}_{o}\rightarrow \mathbf{x}(t)=T_{t_{o},t}\mathbf{x}_{o}
\label{flow}
\end{equation}%
generated by an initial value problem of the type
\begin{equation}
\left\{
\begin{array}{c}
\frac{d\mathbf{x}}{dt}=\mathbf{X}(\mathbf{x},t;Z), \\
\mathbf{x}(t_{o})=\mathbf{x}_{o}.%
\end{array}%
\right.  \label{initial-value problem}
\end{equation}%
Here the notation is standard \cite{Tessarotto2004,Ellero2005}. Thus, $%
T_{t_{o},t}$ is the evolution operator generated by
\begin{equation}
\mathbf{X}(\mathbf{x},t;Z)=\left\{ \mathbf{v,F}\right\} \mathbf{,}
\label{vector-field X}
\end{equation}%
$\chi (\mathbf{x}_{o},t_{o},t)$ being the solution of the initial-value
problem and $\mathbf{F}(\mathbf{x},t;Z)$\ a vector field (denoted as \textit{%
mean-field force}) to be suitably prescribed. The definitions of the PDF and
of its functional class $\left\{ f_{1}\right\} $ depend on the type of
relationship established between the fluid fields and the PDF$,$ to be
prescribed in some suitable sense$.$\ In CSM-inspired statistical models
this is realized by means of a PDF\textit{, }$f_{1},$ which is assumed to
satisfy the corresponding $1-$\textit{point Liouville equation -} here
denoted as \textit{inverse kinetic equation} (IKE \cite%
{Tessarotto2004,Ellero2005}) - of the form
\begin{equation}
L(\mathbf{r,v},t;f_{1})f_{1}(\mathbf{r,u,}t;f_{1})=0.  \label{Liouvlle eq}
\end{equation}%
Here $L(\mathbf{r,v},t;f_{1})$ denotes the Liouville streaming operator
\begin{eqnarray}
&&\left. L(\mathbf{r,v},t;f_{1})\cdot \equiv \frac{\partial }{\partial t}%
\cdot +\frac{\partial }{\partial \mathbf{x}}\cdot \left\{ \mathbf{X}(\mathbf{%
x},t;Z)\cdot \right\} \equiv \right.  \label{Liouville op} \\
&&\left. \equiv \frac{\partial }{\partial t}\cdot +\mathbf{v\cdot }\frac{%
\partial }{\partial \mathbf{r}}\cdot +\frac{\partial }{\partial \mathbf{v}}%
\cdot \left\{ \mathbf{F}(\mathbf{x},t;f_{1})\cdot \right\} \right. ,  \notag
\end{eqnarray}%
while the vector field $\mathbf{F}$ is generally to be assumed functionally
dependent on $f_{1}.$ In particular, in the HRE and ML approaches this is
obtained by invoking the position
\begin{equation}
\mathbf{F\equiv F}_{H},  \label{HRE positiion}
\end{equation}%
with $\mathbf{F}_{H}$ denoting the total fluid force density acting on each
fluid element [see Eq.(\ref{2c}) in the Appendix] and identifying $f_{1}$
with the particular solution
\begin{equation}
f_{1}=f_{H}  \label{Dirac detla}
\end{equation}%
[with $f_{H}$ defined by Eq.(\ref{DIRAC DELTA- HRE})]. Hence in this case $%
\left\{ f_{1}\right\} $ is manifestly the functional class of distributions
of the form (\ref{DIRAC DELTA- HRE}).

The HRE and ML approaches, both characterized by the same statistical model $%
\left\{ f_{H},\Gamma \right\} $ (\textit{HRE-ML statistical model}), belong
actually to a more general class of statistical models inspired by Classical
Statistical Mechanics (CSM) (see also Sec.2). \

In particular, this concerns the representation of all the relevant
dynamical variables in terms of \textit{hidden variables }\cite%
{Tessarotto2008-1,Tessarotto2009} (see Appendix A). By definition they
denote a suitable set of independent variables $\mathbf{\alpha }=\left\{
\alpha _{i},i=1,k\right\} \in V_{\mathbf{\alpha }}\subseteq \mathbf{R}^{k},$
with $k\geq 1,$ which cannot be known deterministically, i.e., are not
observable. In the context of turbulence theory these variables are
necessarily stochastic. This means that they are characterized by a suitable
\textit{stochastic probability density} $g$ defined on $V_{\mathbf{\alpha }}$
(see definitions and related discussion in Appendix A, Subsection A.2),
while the ensemble average $\left\langle \cdot \right\rangle $ [defined in
Eq.(\ref{f_1-HRE APPROACH})] can be identified with the stochastic-averaging
$\left\langle \cdot \right\rangle _{\mathbf{\alpha }}$ defined by Eq.(\ref%
{stochastic averaging operator}) [see Appendix A]. \ Hence, for turbulent
flows the fluid fields - together with the PDF $f_{1}$ and the vector field $%
\mathbf{F}(\mathbf{x},t;Z)$ appearing in Eqs.(\ref{vector-field X}) and (\ref%
{Liouvlle eq}) - admit a\textit{\ }representation of the form \cite%
{Tessarotto2008-1,Tessarotto2009}

\begin{equation}
\left\{
\begin{array}{c}
\left\{ Z\right\} =\left\{ Z(\mathbf{r},t,\mathbf{\alpha })\right\} \\
f_{1}=f_{1}(\mathbf{r,u,}t,\mathbf{\alpha ;}Z) \\
\mathbf{F}=\mathbf{F}(\mathbf{x},t,\mathbf{\alpha };Z)%
\end{array}%
\right.  \label{stochastic functions}
\end{equation}%
to be defined in terms of a set of \ hidden variables $\mathbf{\alpha }$ and
a stochastic model $\left\{ g,V_{\mathbf{\alpha }}\right\} $ (see again
Appendix A, Subsection A.2). Hence, $\left\{ Z\right\} ,$ $f_{1}$ and $%
\mathbf{F}(\mathbf{x},t;Z)$ are necessarily \textit{non-observable}.
Nevertheless, if we assume that the fluid fields $\left\{ Z\right\} $ are
uniquely-prescribed ordinary functions of $(\mathbf{x,}t,\mathbf{\alpha })$
defined for all $(\mathbf{x,}t,\mathbf{\alpha })\in \overline{\Gamma }\times
I\times V_{\alpha },$ it follows that they can still be considered \textit{%
conditional observables }(see Appendix A, Subsection A.1). Similar
conclusions apply to $f_{1},$ and to the vector field $\mathbf{F}(\mathbf{x}%
,t;Z)$ as well$.$

\subsection{1B - The Closure problem}

Based on these requirements [the positions defined by Eqs.(\ref{HRE
positiion}) and (\ref{Dirac detla})], in turbulence theory, i.e., when the
fluid fields $\left\{ Z\right\} $ are considered as stochastic functions,
usually the statistical description involves the construction of an infinite
set of \textit{continuous }many-point PDF's which obey a hierarchy of
statistical equations, the so-called ML (Monin-Lundgren \cite%
{Monin1967,Lundgren1967}) hierarchy. Therefore in this case the statistical
model actually involves the identification of $\left\{ f\right\} $ with the
functional class of the many-point PDF's$.$ As proven by Hosokawa \cite%
{Hosokawa2006}, the problem can be formulated in an equivalent way in the
framework of the HRE approach, yielding the well-known Hopf\textit{\ }$\phi $%
\textit{\ }functional-differential\ equation.

To date, the search of possible exact "closure conditions" for the ML
hierarchy (or \textit{closure problem}) - or equivalent, of exact solutions
of the HRE approach - remains one of the major unsolved theoretical problems
in fluid dynamics. For the ML-approach, this involves in principle the
construction of statistical models which should be characterized by a finite
number of (multi-point) PDF's, i.e., determined in such a way that the time
evolution of the fluid fields can be uniquely determined in terms of them.
In practice the program of constructing theories of this type, and holding
for arbitrary fluid fields, is still open due to the difficulty of
preserving the full consistency with the fluid equations. In fact, it is
well known that many of the customary statistical models adopted in
turbulence theory - which are based on closure conditions of various type
(see for example Monin and Yaglom \cite{Monin1975} 1975 and Pope, 2000 \cite%
{Pope2000}) - typically reproduce at most only in some approximate (i.e.,
asymptotic) sense the fluid equations.

In particular, an interesting related issue is that posed by the
determination of the form of the $1-$point PDF. In decaying isotropic
turbulence, according to some authors (see in particular Batchelor \cite%
{Batchelor}) this is predicted as \textit{almost-Gaussian}. \ Although
others, based on the adoption of suitable model dynamical systems for NS
turbulence (Falkovich and Lebedev \cite{Falkovich1997} and Li and Menevau
\cite{Li 2005}), have pointed that the tails of the 1-point PDF might
exhibit a strongly \textit{non-Gaussian behavior}, according to more recent
investigations (Hosokawa \cite{Hosokawa2008}) there seems to be still
insufficient experimental evidence for a generalized behavior of this type,
at least in the case of homogeneous turbulence.

The basic difficulty is, however, related to the proper formulation of a
rigorous theory for the 1-point PDF holding for arbitrary NS fluids. In this
reference, in particular, an important issue is related to the quest of a
possible \textit{exact statistical evolution equation} for the $1-$point PDF
\cite{Tessarotto2009}, \textit{holding both for deterministic and stochastic
fluid equations}, which is capable of yielding the correct fluid fields and
holds for arbitrary initial and boundary conditions of the relevant related
physical observables (or, respectively, conditional observables). \

This refers, in particular, to the subset of statistical models $\left\{
f,\Gamma \right\} $ in which the PDF $f$ \ is considered as an ordinary
function.

An example is provided by IKT \cite{Tessarotto2004,Ellero2005} in which the $%
1-$point PDF obeys by construction, both for regular fluids (i.e.,
deterministic) and turbulent (i.e., stochastic) fluid fields, an IKE of the
form (\ref{Liouvlle eq}).

This result raises the interesting question whether, in some suitable
setting, i.e., \textit{for appropriate statistical models }and in particular
in the case in which\textit{\ }$\left\{ f\right\} $ is a set of\textit{\ }%
ordinary functions, the closure problem can actually be solved. In this
regard, there are actually two possible routes which seem currently
available:

\begin{itemize}
\item \textit{the first route}: for a prescribed statistical model $\left\{
f,\Gamma \right\} $ it involves the search of possible finite subset of
statistical equations (formed by the $s$ equations for the PDF's, $f_{n},$
having $n\leq s$ and $s$ finite) which define a closed set of\ equations$;$

\item \textit{the second route}: is based on the search of possible
alternative statistical models $\left\{ f,\Gamma \right\} ,$ based on the
hidden-variable representation (\ref{stochastic functions}) and \textit{%
mathematically equivalent to the complete set of fluid equations for INSF,
for which }$\Gamma $ \textit{coincides with the restricted phase-space and
the PDF, identified with the }$1-$\textit{point (or local) PDF } $f_{1}$,
\textit{satisfies an IKE of the type defined by (\ref{Liouvlle eq}), in
which the mean-field force }$\mathbf{F}$ \textit{depends functionally only
from }$f_{1},$ via suitable moments of the PDF$.$ \textit{Statistical models
of this type (in which the evolution operator depends only on }$f_{1},$%
\textit{\ via its moments) are usually said to satisfy a kinetic closure
condition}. $\ $
\end{itemize}

The first approach, by far the most popular one in the literature and
adopted by several authors for the construction of approximate closure
models of the ML hierarchy, poses nevertheless - as indicated above - a
problem of formidable difficulty.

Instead, a possible candidate for statistical models of the second type,
adopting the hidden-variable representation (\ref{stochastic functions}), is
already known \cite{Tessarotto2008-1,Tessarotto2009}. It is provided by the $%
1-$point PDF $f_{1}$ which characterizes the IKT for an incompressible NS
fluid \cite{Tessarotto2004,Ellero2005}. In this case the PDF can always be
required to be also a \textit{velocity-space probability density}, i.e. to
satisfy the normalization $\int\limits_{U}d^{3}\mathbf{v}$ $f_{1}=1.$

In both cases, however, it must be stressed that 'a priori' the definition
of relevant statistical model still remains essentially arbitrary. This
concerns, besides the choice of the phase-space $\Gamma $ (only for the
first route) and of the functional class $\left\{ f_{1}\right\} $ to which
the PDF belongs, also the definition of the set of moments to be associated
to the fluid fields $\left\{ Z\right\} $. In particular, in IKT in principle
the definition of $f_{1}$ is non-unique because higher-order moments of $%
f_{1}$ may be still undetermined. \ In addition $f_{1}$ - even if assumed as
an ordinary function - still remains by definition \textit{non-observable}.

\subsection{1C - Physical realizability conditions on the $1-$point PDF}

In reference to statistical models based on $f_{1},$such as the IKT model, a
natural question arises, i.e., whether the arbitrariness in their definition
can be used,\textit{\ by proper prescription on its functional class} $%
\left\{ f_{1}\right\} ,$\textit{\ to determine it uniquely consistent not
only with INSE but also with the relevant physical observables }(or
conditional observables)\textit{. }In such a case $f_{1}$ might, in
particular, be viewed as a \textit{conditional observable }too\textit{. }

An important preliminary task to accomplish is to establish \ the
relationship of $f_{1}$ with the fluid fields. More precisely, here we wish:

A) \textit{to assess whether - besides the complete set of fluid fields -
the PDF is possibly related to additional observables or conditional
observables. This information, in fact, might provide an effective
constraint on the functional class of the initial PDF - evaluated at the
initial time (}$t=t_{o}$\textit{) - }$\left\{ f_{1}\right\} _{r_{o}};$

B) \textit{to determine whether, for suitable fluid fields, there exists a
deterministic limit for }$f_{1},$ \textit{namely for which it is a Dirac
delta of the type (\ref{Dirac detla}). Such a limit would manifestly provide
a constraint on the functional class }$\left\{ f_{1}\right\} .$

Let us point out that for both problems a simple solution actually exists.

\subsubsection{1C.a - The $1-$\textit{point velocity-frequency density}
\textit{function}}

In particular, regarding the observable, here we claim that it can be
identified with the \textit{configuration-space average} of $1-$point PDF, $%
f_{1}(t)\equiv f_{1}(\mathbf{r,v,}t,\mathbf{\alpha };Z)$, namely $%
\left\langle f_{1}(t)\right\rangle _{\Omega },$ $\left\langle \cdot
\right\rangle _{\Omega }$ denoting the $\Omega -$averaging operator. For a
generic phase function $a(\mathbf{r,v,}t,\mathbf{\alpha })$ its
configuration-space average on $\Omega $, can be\ identified respectively
either with the continuous or discrete operators
\begin{equation}
\left\langle a(\mathbf{r,v,}t,\mathbf{\alpha })\right\rangle _{\Omega
}\equiv \frac{1}{\mu (\Omega )}\int\limits_{\Omega }d^{3}\mathbf{r}a(\mathbf{%
r,v,}t,\mathbf{\alpha })  \label{f_1-average}
\end{equation}%
\begin{equation}
\left\langle a(\mathbf{r,v,}t,\mathbf{\alpha })\right\rangle _{\Omega
}\equiv \lim_{N\rightarrow \infty }\frac{1}{N}\sum\limits_{i=1,N}a(\mathbf{%
r,v,}t,\mathbf{\alpha }).  \label{f_1 average-discrete}
\end{equation}%
In practice, for actual comparisons with experimental data, both operators
can be conveniently replaced by a finite summation of the form
\begin{equation}
\left\langle a(\mathbf{r,v,}t,\mathbf{\alpha })\right\rangle _{\Omega }\cong
\overline{a}^{(A)}(\mathbf{v,}t,\mathbf{\alpha })\equiv \frac{1}{N_{\ast }}%
\sum\limits_{i=1,N_{\ast }}a(\mathbf{r}_{i}\mathbf{,v,}t,\mathbf{\alpha }),
\label{f_1 discrete- approximate}
\end{equation}%
$N_{\ast }$ denoting a suitable integer to be considered $\gg 1$. \ Thus, $%
\left\langle f_{1}(t)\right\rangle _{\Omega }$ should be identified with the%
\textit{\ }$1-$\textit{point velocity-frequency density} \textit{function}
(VFDF). Namely, introducing the short- way notation $\widehat{f}%
_{1}^{(freq)}(t)\equiv \widehat{f}_{1}^{(freq)}(\mathbf{r}_{i}\mathbf{,v,}t,%
\mathbf{\alpha }),$ it should result
\begin{equation}
\left\langle f_{1}(t)\right\rangle _{\Omega }=\widehat{f}_{1}^{(freq)}(t).
\label{constraint on f1 (at initial time)}
\end{equation}%
where
\begin{equation}
\widehat{f}_{1}^{(freq)}(t)\equiv \lim_{N\rightarrow \infty }\frac{1}{N}%
\sum\limits_{i=1,N}N_{1}(\mathbf{r}_{i}\mathbf{,v,}t,\mathbf{\alpha };Z)
\label{stat-1}
\end{equation}%
and $N_{1}(\mathbf{r}_{i}\mathbf{,v,}t,\mathbf{\alpha };Z)$ is the frequency
associated to the fluid velocity field $\mathbf{V}(\mathbf{r,}t,\mathbf{%
\alpha })$ [see related discussion in Appendix B and in particular Eqs. (\ref%
{B.3})-(\ref{B.6})]. Hence, by definition, from Eq.(\ref{constraint on f1
(at initial time)}) it follows that it must be%
\begin{equation}
\int\limits_{U}d^{3}\mathbf{v}\left\langle f_{1}(t)\right\rangle _{\Omega
}=\int\limits_{U}d^{3}\mathbf{v}\widehat{f}_{1}^{(freq)}(t)=1.
\label{stat-2}
\end{equation}

\subsection{1C.b - The deterministic limit of $f_{1}$}

Regarding the search of a deterministic limit for $f_{1},$\ there is
manifestly only one case when this can happen and it occurs in the limiting
case of the stationary solution (of INSE) $Z_{o}\equiv \left\{ \mathbf{V=0}%
,p_{1}=0\right\} $ (or \textit{null solution}), for which $f_{1}$ $\equiv
\left\langle f_{1}\right\rangle _{\Omega }.$ In fact, in such a case $f_{1}$
manifestly must coincide with a distribution, i.e.,
\begin{equation}
\lim_{\substack{ p_{1}\rightarrow 0^{+}  \\ \left\vert \mathbf{V}\right\vert
\rightarrow 0^{+}}}f_{1}=\delta (\mathbf{v})  \label{Dirac Delta}
\end{equation}%
(\textit{deterministic limit}), while also there manifestly results%
\begin{equation}
\lim_{\substack{ p_{1}\rightarrow 0^{+}  \\ \left\vert \mathbf{V}\right\vert
\rightarrow 0^{+}}}\widehat{f}_{1}^{(freq)}(t)=\delta (\mathbf{v}).
\label{Dirac delta-2}
\end{equation}

Since the null solution can always in principle be reached (i.e., moving
backward in time and by application of a combination of appropriate volume
forces and boundary conditions acting on the fluid, both to be suitably
defined), it follows that Eq.(\ref{Dirac Delta}) might always be regarded as
\textit{a possible alternative initial condition} for $f_{1}.$ In the
following we intend to prove that, as an example, it can be represented as
the limit function
\begin{equation}
\delta (\mathbf{v})\equiv \lim_{\substack{ p_{1}\rightarrow 0^{+}  \\ %
\left\vert \mathbf{V}\right\vert \rightarrow 0^{+}}}f_{M}(\mathbf{v};p_{1})
\label{limit function}
\end{equation}%
of a \textit{Gaussian PDF }
\begin{equation}
f_{M}(\mathbf{v-V};p_{1})=\frac{1}{\pi ^{2}v_{th,p}^{3}}\exp \left\{ -\frac{%
\left\Vert \mathbf{v-V}\right\Vert ^{2}}{v_{thp}^{2}}\right\} ,
\label{Maxwellian PDF}
\end{equation}%
where $\mathbf{V=0}$ and $v_{thp}=\left( 2p_{1}/\rho _{o}\right) ^{1/2}$ is
the thermal velocity due to $p_{1}$ [see Eq.(\ref{kinetic pressure})]$.$ It
is obvious, however, that the previous requirement does not provide a unique
functional form for $f_{1}.$ For instance, when both $\mathbf{V}(\mathbf{r}%
,t,\mathbf{\alpha })$ and $p_{1}(\mathbf{r},t,\mathbf{\alpha })$ are
considered infinitesimals (of order $\varepsilon $), Eq.(\ref{Dirac Delta})
only requires that
\begin{equation}
f_{1}(t)=f_{M}(\mathbf{u};p_{1},\mathbf{\alpha })+\delta f_{1}(\mathbf{r,u,}%
t,\mathbf{\alpha };Z),
\end{equation}%
with $\delta f_{1}$ infinitesimal of order $O(\varepsilon ).$

In the following $\left\{ f_{1},\Gamma \right\} $ will be denoted as \textit{%
statistically complete} if the PDF $f_{1}$ :

A) admits for all $\left( \mathbf{r,}t\right) \in \overline{\Omega }\times I$
\ (including the initial time $t_{o}$) and $G=1,\mathbf{v,}u^{2}/2,\mathbf{uu%
},\mathbf{u}u^{2}/2$ the velocity and phase-space moments $\int\limits_{U}d%
\mathbf{v}Gf_{1}$ and $\int\limits_{\Gamma }d\mathbf{v}f_{1}\ln f_{1}$ and
satisfies the constraint equations%
\begin{eqnarray}
\int\limits_{U}d\mathbf{v}Gf_{1} &=&1,\mathbf{V}(\mathbf{r,}t,\mathbf{\alpha
}),p_{1}(\mathbf{r,}t,\mathbf{\alpha }),  \label{MOMENTS} \\
S(f_{1}(t)) &=&S_{T}.  \label{ENTROPY}
\end{eqnarray}%
Here $p_{1}(\mathbf{r,}t,\mathbf{\alpha })>0$ denotes the \textit{kinetic
pressure}
\begin{equation}
p_{1}(\mathbf{r},t,\mathbf{\alpha })=p(\mathbf{r},t,\mathbf{\alpha }%
)+p_{0}(t,\mathbf{\alpha })-\phi (\mathbf{r},t,\mathbf{\alpha }),
\label{kinetic pressure}
\end{equation}%
with $p_{0}(t,\mathbf{\alpha })>0$ (the \textit{pseudo-pressure}) a strictly
positive, smooth, real function and $\phi (\mathbf{r},t,\mathbf{\alpha })$ a
suitably defined potential;

B) at the initial time $t=t_{o}$ satisfies the constraint (\ref{constraint
on f1 (at initial time)});

C) satisfies the constraint defined by the deterministic limit (\ref{Dirac
Delta}).

The constraint equations (\ref{MOMENTS})-(\ref{kinetic pressure}),(\ref%
{constraint on f1 (at initial time)}) and (\ref{Dirac Delta}) are here
denoted as \textit{physical realizability conditions}.

\subsection{1D - Open issues}

Here we shall consider a class of statistical models $\left\{ f_{1},\Gamma
\right\} ,$ based on the $1-$point PDF, $f_{1}$ [with $f_{1}$ defined on $%
\Gamma $ and $\Gamma $ identified with (\ref{restricted phase space})],
which yield a \textit{complete inverse kinetic theory for the INSE problem}$%
. $ In other words, each $\left\{ f_{1},\Gamma \right\} $ should yield the
\textit{complete set of fluid fields},$\left\{ Z\right\} $, \ to be\textit{\
represented in terms of suitable velocity moments of the same} PDF. In
addition, $\left\{ f_{1},\Gamma \right\} $ will be required to hold for
arbitrary fluid fields $\left\{ Z\right\} $ which, in the domain of
existence $\overline{\Omega }\times I,$ are\textit{\ }strong solutions of
the INSE problem, the latter - for greater generality - to be considered
\textit{either deterministic or stochastic} (see Appendix A).

In the construction of statistical models of this type several interesting
issues arise, which are related both to the prescription of the initial
conditions on the PDF and to its time evolution. In particular, they concern
whether there exists a statistical model $\left\{ f_{1},\Gamma \right\} ,$
such that:

\begin{enumerate}
\item (\textit{Problem 1: uniqueness condition}) both $\left\{ f_{1},\Gamma
\right\} $ and the PDF $f_{1}$ are \textit{unique; }

\item (\textit{Problem 2: conditions of statistical completeness}) \ $%
\left\{ f_{1},\Gamma \right\} $ is statistically complete, namely $f_{1}$
satisfies the requirements posed in Subsection 1C\textit{. }

Here we shall consider, in particular, the case in which the frequency $%
\widehat{f}_{1}^{(freq)}(t_{o})$ is an ordinary function (i.e., not a
distribution), consistent with the requirement that $f_{1}$ is an ordinary
function too. The second requirement is that in the limit in which $%
p_{1}\rightarrow 0^{+}$ and $\left\vert \mathbf{V}\right\vert \rightarrow
0^{+},$ there results (\ref{Dirac Delta}).

\item (\textit{Problem 3: moment-closure condition}) can be defined in such
a way that it satisfies a \textit{moment-closure condition.\ }In other
words: \ whether there exists a finite set of moment equations of IKE (\ref%
{Liouvlle eq}) which are closed. This is actually a basic requirement of
IKT. Hence, it\ should be satisfied by construction, if $\left\{
f_{1},\Gamma \right\} $ relies on IKT \cite{Ellero2005}.

\item (\textit{Problem 4: kinetic-closure condition}) it can be defined in
such a way that at any time $t\geq t_{o}$ (with $t\in I$), the statistical
equation advancing in time $f_{1}$ \textit{depends, besides }$f_{1},$\textit{%
\ }only \textit{on a finite number of moments of the same PDF}$,$ which
include necessarily \textit{the complete set of fluid fields} $\left\{ Z(%
\mathbf{r,}t)\right\} .$ In such a case $\left\{ f_{1},\Gamma \right\} $ is
said to satisfy a \textit{kinetic closure condition.} This assumption is -
in some sense - analogous to the closure problem for the ML hierarchy. In
both cases it effectively involves the construction of the mean-field force
which advances in time the $1-$point PDF.

\item (\textit{Problem 5: determination of multi-point PDF's}) \textit{the
statistical model }$\left\{ f_{1},\Gamma \right\} $ \textit{can be
constructed in such a way that determines uniquely the multi-point PDF's, as
well as the related observables. }

\item (\textit{Problem 6: closure condition of the statistical equations for
multi-point PDF's}) the statistical equations for the multi-point PDF's
depend only on $f_{1}.$
\end{enumerate}

In the remainder a statistical model which satisfies both the moment and
kinetic closure conditions indicated above in Problems 3 and 4 will be
denoted as \textit{closed}. \textit{\ }

\subsection{1E - Goals and scheme of presentation}

Here we claim that the IKT statistical model can be defined in such a way to
provide an explicit solution of problems 1-6. The construction of the $1-$%
point PDF is achieved adopting the so-called inverse kinetic theory (IKT)
for fluid dynamics (Tessarotto \textit{et al.}, 2004-2009 \cite%
{Tessarotto2004,Ellero2005,Tessarotto2008-1,Tessarotto2009,Tessarotto2006,Tessarotto2007a,Tessarotto2007b,Tessarotto2008-2,Tessarotto2008-4,Tessarotto2008-5,Tessarotto2008-6,Tessarotto2008-7,Tessarotto2008-3,Tessarotto2008-8,Tessarotto2009b}
).\ This type of approach can be formulated both when the fluid is
considered \textit{regular} and \textit{turbulent, i.e. }the corresponding
fluid equations (INSE) and the fluid fields $\left\{ Z\right\} $ are
respectively considered \textit{deterministic} and \textit{stochastic }\cite%
{Tessarotto2008-1,Tessarotto2009}\textit{. }In particular, $f_{1}$ is
defined in such a way that:

A) both the PDF and the related mean-field force ($\mathbf{F}$) which
determines its time evolution [see Eq.(\ref{vector-field X})] are \textit{%
conditional observables. }This concept [see Appendix A, Subsection A.1]\ is
shown to imply the uniqueness and closure properties of the statistical
model;

B) the PDF satisfies suitable physical realizability conditions. This
requires, in particular, that the complete set of fluid fields\textit{\ }$%
\left\{ Z\right\} $ is necessarily represented in terms of moments the PDF%
\textit{\ }\cite{Tessarotto2004,Ellero2005,Tessarotto2008-1}, while the
initial PDF must be defined so that its configuration-space average is
suitably prescribed [see Eq.(\ref{constraint on f1 (at initial time)})]. In
such a case, the solution of the initial condition for $f_{1}$ (see Problem
2) is uniquely achieved by invoking the Principle of Entropy Maximization
(PEM; Jaynes, 1957 \cite{Jaynes1957}). This permits to identify two possible
solutions for the initial PDF, corresponding respectively to the initial
condition 1$_{A}$ and to 1$_{A}$ and 1$_{B}$ jointly.

The first one is realized by a local Gaussian PDF ($f_{M}$).

The formulation of the corresponding statistical model $\left\{ f_{M},\Gamma
\right\} $ and the analysis of its basic properties are discussed in Section
2. First, it is pointed out that PEM requires necessarily that $f_{1}\equiv
f_{M}$ must be a conditional observable for all $t\in I$. \ Then, by
identifying the mean-field force $\mathbf{F}(f_{M})$\ with a conditional
observable, it is proven that it is uniquely defined, both as a function of
the kinetic velocity $\mathbf{v}$ (actually a polynomial of second degree
with respect the relative kinetic velocity $\mathbf{u=v-V}$)\textbf{\ }and
of the fluid fields $\left\{ Z\right\} \equiv \left\{ \mathbf{V}%
,p_{1}\right\} .$ As a consequence, the statistical model $\left\{
f_{M},\Gamma \right\} $ is unique (THM.1; see also Problem 1). Moreover, the
complete set of fluid fields $\left\{ Z\right\} $ are uniquely determined as
moments of $f_{M},$ which means that the classical dynamical system defined
by the initial-value problem (\ref{initial-value problem}) determines
uniquely both the time-evolution of the PDF and of the complete set of fluid
fields $\left\{ Z\right\} $. However, unless the initial constraint (\ref%
{constraint on f1 (at initial time)}) is fulfilled by $f_{M}$ (which is
generally not the case)$,$ $\left\{ f_{M},\Gamma \right\} $ is not
statistically complete (see Corollary to THM.1)$.$

The construction of the statistical model $\left\{ f_{1},\Gamma \right\} $
which takes into account such a case is carried out in Section. 3. The
condition of statistical completeness requires that the initial PDF
satisfied simultaneously conditions 1$_{A}$ and 1$_{B}.$ It is found that
this generally provides (at $t=t_{o}$) a non-Gaussian initial PDF of
prescribed form (see THM.2). Its time evolution depends now on the
mean-field force $\mathbf{F}(f_{1}).$ Under the assumptions that its
dependence in terms of the kinetic velocity $\mathbf{v}$ can only occur, as
in the Gaussian PDF, only via a polynomial of second degree in $\mathbf{u}$
and identifying $\mathbf{F}(f_{1})$, as in the previous case, with a
conditional observable its form is found to be unique as the statistical
model $\left\{ f_{1},\Gamma \right\} $ $\ $while $\left\{ f_{1},\Gamma
\right\} $ is also statistically complete.

In the same section (Subsection 3D) the closure problem (Problems 3 and 4),
based on the IKT statistical model, is formulated. This refers, in
particular, to the construction of its formal solution (i.e., see also
paragraphs 1B and 1D). This is achieved, \textit{both for deterministic and
stochastic fluids described by the INSE problem }(as defined in Appendix A).
In our theory this is done by determining directly $f_{1}$ , rather than its
stochastic average $\left\langle f_{1}\right\rangle .$ As a basic
consequence, it follows that the statistical equation for $f_{1}$ is
necessarily closed, namely it depends only on $f_{1}$ and a finite number of
moments of the same PDF. The proof of the closure property of the IKT
statistical model is proven in THM.3.

There are several new contributions and basic consequences of the theory
here presented.

Besides the (generally non-Gaussian) characterization of the initial PDF, it
is found that, both for regular and turbulent flows, its time evolution is
provided by a Liouville equation, while the corresponding Liouville operator
depends only on a finite number of velocity moments of the same PDF. Hence,
its time evolution depends (functionally) solely on the same (1-point) PDF.
An interesting issue is also provided by the comparison, carried out in
section 4, between $\left\{ f_{1},\Gamma \right\} $ and the common
statistical model, denoted $\left\{ f_{H},\Gamma \right\} ,$ laying at the
core of the customary statistical approaches [i.e., the HRE \cite{Hopf1952,
Rosen1960,Edwards1964} and ML \cite{Monin1967,Lundgren1967} approaches]. The
latter, although both unique and closed (in the sense of Problems 3 and 4),
is proven to be\textit{\ statistically incomplete} (see THM.4)\textit{,}
since the corresponding PDF cannot generally fulfill the physical constraint
placed on it by Eq.(\ref{constraint on f1 (at initial time)})\textit{\ }[at%
\textit{\ }$t=t_{o}$ or at any time $t\in I]$. \

The connection with previous statistical approaches is investigated. In
particular, it is shown that $f_{1}$ can be identified with a suitable
stochastic average of the PDF $f_{H}$ (see THM.5). A remarkable consequence
of the present theory is that it affords \textit{the exact construction} of
the corresponding statistical equation for the stochastic-averaged
PDF.(Section 4). The latter is shown to depart from the customary transport
equation considered in the literature [see, for example, Dopazo \cite{Dopazo}
and Pope \cite{Pope2000}]. The relationship with the Hopf
functional-differential method \cite{Hopf1952} is also displayed.

Finally, the explicit construction of multi-point PDF's and of the related
observables (Section 5) is achieved. In particular, the statistical
evolution equations for the multi-point PFD's are shown to maintain the form
of Liouville equations. \ As a practical application, the explicit
construction of reduced $2-$point PDF's - and of their related statistical
equations - both usually investigated in experimental/numerical research in
fluid dynamics, is presented. In fact, despite not being themselves
observables, they are nevertheless related to physical observables (or
conditional observables), such as the velocity difference between different
fluid elements, usually adopted for the statistical analysis of turbulent
fluids. Finally, in Section 6 the conclusions are drawn.

\

\section{2 - IKT approaches}

A fundamental aspect of fluid dynamics is the construction of statistical
models $\left\{ f_{1},\Gamma \right\} $ in which the $1-$point PDF is the
solution of a so-called \textit{inverse problem, }involving the search of a
so-called inverse kinetic theory (IKT) \textit{able to yield the complete
set of fluid equations for the fluid fields. }A particular realization for $%
\left\{ f_{1},\Gamma \right\} $ is provided by (the already mentioned) ML
approach, which is based on the position (\ref{HRE positiion}) and the
particular solution (\ref{Dirac detla}). In such a case it follows, by
construction, that $f_{1}$ depends explicitly, and not just merely in a
functional sense, on the fluid field $\mathbf{V}(\mathbf{r,}t).$ Hence, IKE (%
\ref{Liouvlle eq}) implies necessarily INSE (and therefore can be viewed as
an inverse kinetic equation). Nevertheless, it is obvious that the fluid
pressure $p(\mathbf{r,}t)$ \textit{cannot be represented as a \ moment of
the same PDF. }

In this connection, however, a \ more general viewpoint is represented by
the search of so-called \textit{complete IKT's}\textit{\ }able to yield as
moments of the PDF the \textit{whole set of \ fluid fields }$\left\{
Z\right\} $ which determine the fluid state and in which the same PDF
satisfies a Liouville equation. This implies that in such a case\textit{\
there must exist a classical dynamical system, of the type defined by Eq.(%
\ref{flow}), whose evolution operator }$T_{t_{o},t}$ \textit{advances in
time both the PDF and the related fluid fields, while preserving - at the
same time - a suitable probability measure }(Frisch, 1995 \cite{Frisch1995})%
\textit{. }Despite previous attempts (Vishik and Fursikov, 1988 \cite%
{Vishik1988} and Ruelle, 1989 \cite{Ruelle1989}) the existence of such a
dynamical system has remained for a long time an unsolved problem.

This type of approach has actually been achieved for incompressible NS
fluids \cite{Tessarotto2004,Ellero2005}, with the discovery of the
(corresponding) \textit{NS dynamical system} which advances in time the
complete set of fluid fields $\left\{ Z\right\} .$ Its applications\ and
extensions are wide-ranging and concern in particular: incompressible
thermofluids \cite{Tessarotto2008-2}, quantum hydrodynamic equations (see
\cite{Tessarotto2007a,Tessarotto2008-4}), phase-space Lagrangian dynamics
\cite{Tessarotto2008-5}, tracer-particle dynamics for thermofluids \cite%
{Tessarotto2008-6,Tessarotto2009b}, the evolution of the fluid pressure in
incompressible fluids \cite{Tessarotto2008-3}, turbulence theory in
Navier-Stokes fluids \cite{Tessarotto2008-1,Tessarotto2008-2} and
magnetofluids \cite{Tessarotto2009} and applications of IKT to
lattice-Boltzmann methods \cite{Tessarotto2008-8}. In the following we
intend to investigate, in particular, its consequences for the problems
posed in this paper.

\subsection{2A - The IKT statistical model - Basic assumptions}

Let us now show how a statistical model of this type for the $1-$point PDF
[i.e., $\left\{ f_{1},\Gamma \right\} $], which fulfills the requirements
posed in Problems 1-5 and holds both for regular and stochastic flows \cite%
{Tessarotto2008-1,Tessarotto2009}, can be achieved by suitably modifying the
IKT approach earlier developed by Tessarotto and coworkers \cite%
{Tessarotto2004,Ellero2005}\textit{\ }(see also\textit{\ }Ref.\cite%
{Tessarotto2008-1}).\

Such a theory, it must be stressed, is based on some of the axioms (such as
the conservation of entropy, the principle of entropy maximization or the
regularity assumptions) which are typical of CSM. In particular, the IKT of
Refs. \cite{Ellero2005,Tessarotto2007a}) requires that $f_{1}(t)\equiv f_{1}(%
\mathbf{x,}t,\mathbf{\alpha ;}Z)$ is a particular solution of\ the $1-$point
IKE (\ref{Liouvlle eq}) and satisfies the following assumptions (Axioms
\#0-\#4), imposing that for all $(\mathbf{x,}t)\in \overline{\Gamma }\times
I:$

\begin{itemize}
\item (\textit{Axiom \#0:} regularity) $f_{1}(t)$ is an ordinary, strictly
positive function. By assumption it is measurable, i.e., it admits \ the
velocity- and phase-space moments
\begin{equation}
\int\limits_{U}d\mathbf{v}Gf_{1}  \label{momenrs}
\end{equation}%
respectively for $G=1\mathbf{,v,\rho }_{o}u^{2}/3,\mathbf{uu},\mathbf{u}%
u^{2}/3$ and for $S(f_{1}(t))=-\int\limits_{\Gamma }d\mathbf{x}f_{1}(\mathbf{%
r,u,}t,\mathbf{\alpha };Z)\ln f_{1}(\mathbf{r,u,}t,\mathbf{\alpha };Z),$ the
so-called Boltzmann-Shannon (BS) entropy. \ Moreover, $f_{1}$ is suitably
smooth, in the sense that it is is continuous $\overline{\Gamma }\times I,$
is a particular solution of IKE (\ref{Liouvlle eq}) and its velocity moments
$G_{2},G_{3}\equiv \mathbf{v,\rho }_{o}u^{2}/3$ are respectively of class
\end{itemize}

\begin{equation}
\left\{
\begin{array}{c}
G_{2}\mathbf{\in }C^{(3,1)}(\Omega \times I), \\
G_{3}\mathbf{\in }C^{(2,0)}(\Omega \times I);%
\end{array}%
\right.  \label{NSF functional setting-1}
\end{equation}

\begin{itemize}
\item (\textit{Axiom \#1:} \textit{principle of correspondence}) $f_{1}(t)$
satisfies identically in $\overline{\Omega }\times I$ the constraints (\ref%
{MOMENTS})-(\ref{kinetic pressure}) \ and the fluid fields (including the
kinetic pressure $p_{1})$ can be considered (conditional) observables. As a
part of the same axiom it is required, furthermore, that $f_{1}(\mathbf{r,u,}%
t;Z)$ satisfies suitable kinetic initial and boundary conditions [see Ref.].
The latter are defined so that, by construction, the moments (\ref{MOMENTS}%
)-(\ref{kinetic pressure}) identically satisfy the initial and boundary
conditions prescribed by the INSE problem [see Eqs.(\ref{1ca}) and (\ref{1c}%
) in Appendix A];

\item (\textit{Axiom \#2: moment-closure condition)} the moments equations,
corresponding to $G=1\mathbf{,v,\rho }_{o}u^{2}/3$ and evaluated in terms of
IKE (\ref{Liouvlle eq}), define a system of closed equations, which coincide
with the complete set of fluid equations \ provided by INSE [see Eqs.(\ref%
{1b})-(\ref{1bbb}) in Appendix A];

\item (\textit{Axiom \#3: principle of conservation of entropy, or constant
H-theorem}) $f_{1}(t)$ satisfies the constraint equation [also known as
constant H-theorem \cite{Tessarotto2007a}]
\begin{equation}
\frac{\partial }{\partial t}S(f_{1}(t))=0.  \label{constant-H}
\end{equation}

\item (\textit{Axiom \#4: maximum entropy principle}) $f_{1}(t)$ satisfies,
at $t=t_{o},$ the \textit{constrained maximal variational principle} (also
known as principle of entropy maximization or PEM; Jaynes, 1957 \cite%
{Jaynes1957}):\textit{\ \ }%
\begin{equation}
\delta S(f_{1}(t))=0.  \label{A-2}
\end{equation}
\end{itemize}

In this paper we shall impose, furthermore, two new assumptions introduced
to satisfy the previous problems \textit{(Problem 1-7),} requiring that:%
\textit{\ }

\begin{itemize}
\item (\textit{Assumption \#1}: \textit{conditional observables}) $f_{1}$
and the mean-field force $\mathbf{F}$ are \textit{conditional observables};

\item (\textit{Assumption \#2}: \textit{statistical completeness}) $\left\{
f_{1},\Gamma \right\} $ is statistically complete, namely $f_{1},$ satisfies
the \textit{physical realizability conditions} placed by: (Assumption \#2a)
the velocity moments (\ref{MOMENTS})-(\ref{kinetic pressure}); (Assumption
\#2b) the VFDF, i.e., Eq.(\ref{constraint on f1 (at initial time)});
(Assumption \#2c) the deterministic limit (\ref{Dirac Delta}).
\end{itemize}

\textit{In the remainder Axioms \#0-4 will be assumed to hold, together with
Assumption \#1 and \#2, for the statistical model} $\left\{ f_{1},\Gamma
\right\} $.

\subsection{2B - The Gaussian particular solution}

Let us now show that imposing PEM (at $t=t_{o}$) and requiring solely the
specification of its functional class $\left\{ f_{1}\right\} $ - i.e., not
imposing also the validity of the initial constraint equation (\ref%
{constraint on f1 (at initial time)}) for the 1-point PDF - uniquely
determines its initial value $f_{1}(t_{o})\equiv f(\mathbf{x,}t_{o},\mathbf{%
\alpha };Z).$

For definiteness, let us assume that $\ f_{1}(t_{o})$ is an ordinary,
strictly positive function, requiring initially that it \textit{satisfies
only the constraint provided by the fluid fields,} namely Eq.(\ref{1ca})
[Assumption \#2a] and by the deterministic limit (\ref{Dirac Delta})
[Assumption \#2c]. \ Invoking the method of Lagrange multipliers, the PEM
variational principle [(\ref{A-2})] implies that at the initial time $t_{o}$
and for arbitrary variations $\delta f_{1}(\mathbf{r,u,}t,\mathbf{\alpha }%
;Z) $ it must result identically$:$
\begin{eqnarray}
&&\int\limits_{\Gamma }d\mathbf{x\delta }f_{1}(\mathbf{r,u,}t,\mathbf{\alpha
};Z)\left\{ 1+\ln f_{1}(\mathbf{r,u,}t,\mathbf{\alpha };Z)+\right. \\
&&\left. +\lambda _{o}+\mathbf{\lambda }_{1}\cdot \mathbf{u+\lambda }_{2}%
\mathbf{u}^{2}\right\} =0.  \notag
\end{eqnarray}%
Here $\lambda _{o},\mathbf{\lambda }_{1}$\textbf{\ }and\textbf{\ }$\mathbf{%
\lambda }_{2}$ are Lagrange multipliers to be determined by imposing the
correspondence principle [i.e., the moment equations (\ref{MOMENTS})]. It
follows that: 1) at $t=t_{o},$ $f_{1}(\mathbf{r,u,}t,\mathbf{\alpha };Z)$
necessarily coincides with a Gaussian distribution (\ref{Maxwellian PDF})
\cite{Ellero2005,Tessarotto2007a} carrying the fluid velocity $\mathbf{V}(%
\mathbf{r,}t_{o},\mathbf{\alpha })$ and the kinetic pressure $p_{1}(\mathbf{%
r,}t_{o},\mathbf{\alpha });$ 2) Axiom \#0 implies that $p_{1}(\mathbf{r}%
,t_{o})$ must be strictly positive; 3) $f_{M}(t_{o})$ is necessarily a
conditional observable\textit{,} so that when $\mathbf{\alpha }$ and $%
\mathbf{v}$ (or the relative kinetic velocity $\mathbf{u}$) are considered
prescribed, $f_{M}(\mathbf{u\equiv v-V(r,}t_{o},\mathbf{\alpha });p_{1}(%
\mathbf{r},t_{o},\mathbf{\alpha }))$ is unique and depends in a prescribed
way only on observables; \ 4) moreover, due to the arbitrariness of the
choice of the initial time $t_{o}$, the Gaussian PDF (\ref{Maxwellian PDF})
is necessarily a particular solution of IKE (\ref{Liouvlle eq}) \textit{for
all} $t\in I$ ; 5) as a consequence, also the same equation at all times is
a conditional observable in the sense indicated above.

On the other hand,\ imposing for Eq.(\ref{Liouvlle eq}) the particular
solution
\begin{equation}
f_{1}(\mathbf{r,u,}t,\mathbf{\alpha };Z)\equiv f_{M}(\mathbf{u};p_{1}(%
\mathbf{r},t,\mathbf{\alpha }))  \label{Maxwellian case}
\end{equation}%
implies that the vector field $\mathbf{F}$ must depend functionally on $%
f_{M} $ and $\left\{ Z\right\} ,$ i.e., it is of the type $\mathbf{F}%
(f_{M})\equiv \mathbf{F}(\mathbf{x,}t;f_{M})$ \cite%
{Ellero2005,Tessarotto2007a}. One finds, however, that the most general
\textit{admissible form }of the vector field $\mathbf{F}(f_{M})$ [namely one
which is consistent with the requirement (\ref{Maxwellian case})], is of the
type:

\begin{eqnarray}
&&\left. \mathbf{F}(\mathbf{x,}t,\mathbf{\alpha };f_{M})=\mathbf{F}_{0}(%
\mathbf{x,}t,\mathbf{\alpha };f_{M})+\right.  \label{F-tor} \\
&&+\mathbf{F}_{1}(\mathbf{x,}t,\mathbf{\alpha };f_{M})+\Delta \mathbf{F}(%
\mathbf{x,}t,\mathbf{\alpha };f_{M}).  \notag
\end{eqnarray}%
Here $\Delta \mathbf{F}$ is an arbitrary "gauge" vector field satisfying the
homogeneous equation
\begin{equation}
\frac{\partial }{\partial \mathbf{v}}\cdot \left( \Delta \mathbf{F}%
f_{M}\right) =0,  \label{EQUATION}
\end{equation}%
i.e., it does not contribute to IKE when Eq.(\ref{Maxwellian case}) holds
identically, while $\mathbf{F}_{0},$ $\mathbf{F}_{1}$ and $A\mathbf{(r,}%
t;f_{M})$ \ are respectively the vector and scalar fields%
\begin{eqnarray}
&&\left. \mathbf{F}_{0}\mathbf{(x,}t,\mathbf{\alpha };f_{M})=\frac{1}{\rho
_{0}}\mathbf{f}_{R}+\right.  \label{F0} \\
&&\left. +\frac{1}{2}\mathbf{u}\cdot \nabla \mathbf{V}+\frac{1}{2}\nabla
\mathbf{V\cdot \mathbf{u}+}\nu \nabla ^{2}\mathbf{V,}\right.  \notag
\end{eqnarray}%
\begin{eqnarray}
&&\left. \mathbf{F}_{1}\mathbf{(x,}t,\mathbf{\alpha };f_{M})=\frac{1}{2}%
\mathbf{u}A(\mathbf{r,}t;f_{M})+\right.  \label{F1} \\
&&\frac{v_{th}^{2}}{2}\mathbf{\nabla }\ln p_{1}\left\{ \frac{u^{2}}{%
v_{th}^{2}}-\frac{3}{2}\right\} ,  \notag
\end{eqnarray}%
\textit{\ }%
\begin{eqnarray}
&&\left. A\mathbf{(r,}t,\mathbf{\alpha };f_{M})\equiv \frac{1}{p_{1}}\frac{%
\partial }{\partial t}p_{1}-\frac{\rho _{o}}{p_{1}}\left[ \frac{\partial }{%
\partial t}V^{2}/2+\right. \right.  \label{A field} \\
&&\left. +\mathbf{V}\cdot \nabla V^{2}/2-\frac{1}{\rho _{o}}\mathbf{V\cdot f}%
-\nu \mathbf{V\cdot }\nabla ^{2}\mathbf{V}\right] .  \notag
\end{eqnarray}%
Then, by construction $\mathbf{F}_{0}$ and $\mathbf{F}_{1}$ are conditional
observables. In particular Eq.(\ref{EQUATION}) requires%
\begin{equation}
\Delta \mathbf{F}(\mathbf{x,}t,\mathbf{\alpha };f_{M})=\mathbf{u}\cdot
\underline{\underline{\mathbf{E}}}\equiv \Delta \mathbf{F}_{1}(\mathbf{x,}t,%
\mathbf{\alpha };f_{M}),  \label{DEALTA f}
\end{equation}%
where $\underline{\underline{\mathbf{E}}}\equiv \underline{\underline{%
\mathbf{E}}}(\mathbf{r,}t)$ is an arbitrary antisymmetric second-order
tensor.\ From Eq.(\ref{DEALTA f}) it follows that $\Delta \mathbf{F}$ is
manifestly non-observable. In fact, introducing the transformation
\begin{equation}
\Delta \mathbf{F\rightarrow }\Delta \mathbf{F}^{\prime }\mathbf{=}k\Delta
\mathbf{F}  \label{gauge property}
\end{equation}%
with $k\in
\mathbb{R}
$ arbitrary and non-vanishing, it yields for $\mathbf{F}$ an admissible form
too [i.e., consistent with Eq.(\ref{Maxwellian case})].

\subsection{2C - Properties of $\left\{ f_{M},\Gamma \right\} $}

Let us now pose the problem of resolving the indeterminacy of $\mathbf{F}%
(f_{M})$. In Refs. \cite{Tessarotto2007a,Tessarotto2007b} for Gaussian
solutions the uniqueness of $\mathbf{F}$ was achieved based on the
requirement of consistency with extended thermodynamics, namely by imposing
a suitably-prescribed form for higher-order moments of the Liouville
equation. In particular this was found to require
\begin{equation}
\Delta \mathbf{F}\equiv \mathbf{0.}  \label{constraint on DelatF}
\end{equation}

Here we point out, however, that Eq.(\ref{constraint on DelatF}) is
equivalent to impose that\ $\mathbf{F}(\mathbf{x,}t,\mathbf{\alpha };f_{M})$
\textit{is a conditional observable} (see Assumption \#2)\textit{,} namely a
uniquely-prescribed (polynomial) function of the kinetic velocity $\mathbf{v}
$ (as well of the variables and $\mathbf{r,}t,\mathbf{\alpha }$). Hence, the
following theorem holds:

\textbf{THM.1 - Uniqueness of }$\left\{ f_{M},\Gamma \right\} $

\emph{In validity of Assumptions \#1, 2 and 3a,3c the statistical model }$%
\left\{ f_{M},\Gamma \right\} $\emph{\ defined by Eqs.(\ref{Maxwellian PDF}%
), with} \emph{mean-field force} $\mathbf{F}(\mathbf{x,}t,\mathbf{\alpha }%
;f_{M})$ \emph{prescribed by Eqs.(\ref{F-tor})-(\ref{A field}) and subject
to the constraint (\ref{constraint on DelatF}), is unique.\ }

\emph{PROOF}

The proof is immediate. Uniqueness follows, in fact, besides the axioms of
IKT (Axioms \#0-4), from the requirement that both $f_{M}$ and $\mathbf{F}(%
\mathbf{x,}t,\mathbf{\alpha };f_{M})$ be conditional observables and hence,
in particular, from Eq.(\ref{constraint on DelatF}). Q.E.D.

In Ref. \cite{Ellero2005} the statistical model\emph{\ }$\left\{
f_{M},\Gamma \right\} $ was proven to determine uniquely, thanks to Axiom
\#1, the complete set of fluid fields. Such a result holds, however, in
principle for an arbitrary choice of $\Delta \mathbf{F}$ of the form (\ref%
{DEALTA f}), namely also in the case in which $\Delta \mathbf{F\neq 0}$. In
Refs. \cite{Tessarotto2007a,Tessarotto2007b} the uniqueness of the
mean-field force for the Gaussian PDF ($f_{M}$) was achieved based on
phenomenological arguments, i.e., the comparison with extended
thermodynamics. The present result shows, however, that uniqueness tor $%
\mathbf{F}(\mathbf{x,}t,\mathbf{\alpha };f_{M}),$ and hence $\left\{
f_{M},\Gamma \right\} $ too, can actually be achieved based on the physical
prescription that both quantities are conditional observables (see
Assumptions \#1 and 2).

Nevertheless, it is obvious that the statistical model $\left\{ f_{M},\Gamma
\right\} $ generally is \textit{not} \textit{statistically complete,} in the
sense indicated above\textit{. }In fact (generally) $f_{M}$ does not satisfy
the constraint imposed by the initial condition (\ref{constraint on f1 (at
initial time)}. There it follows immediately:\smallskip \medskip

\textbf{COROLLARY to THM.1 - Statistical incompleteness of }$\left\{
f_{M},\Gamma \right\} $

\emph{In validity of THM.1 the statistical model }$\left\{ f_{M},\Gamma
\right\} $\emph{\ is generally statistically incomplete. }

\emph{PROOF}

We notice that for a Gaussian PDF (\ref{Maxwellian PDF}) the deterministic
limit (\ref{Dirac Delta}) exists. This requires letting $p_{1}\rightarrow
0^{+}$ and $\left\vert \mathbf{V}\right\vert \rightarrow 0^{+},$ implying
also that $p_{1}$ must be uniquely determined (i.e., it \ is necessarily an
observable). However, it is obvious the constraint (\ref{1ca}) may not be
fulfilled by $f_{M}.$ In fact, there results generally at $t=t_{o}$ (for
finite $p_{1}$ and $\left\vert \mathbf{V}\right\vert $)%
\begin{equation}
\frac{\widehat{f}_{1}^{(freq)}(\mathbf{v,}t,\mathbf{\alpha };Z)}{%
\left\langle f_{M}\right\rangle _{\Omega }}\neq 1.  \label{completeness}
\end{equation}%
\textit{\ }Q.E.D.

This leaves open the question of extending IKT in such a way to satisfy this
requirement (of statistical completeness). In the remainder we intend to
show that this constraint generally implies a non-Gaussian initial PDF.

\section{3 - IKT statistical model: non-Gaussian case}

An interesting question is posed by the case in which \textit{the PDF} at $%
t=t_{o},$ $f_{1}(t)$ \textit{satisfies also the initial condition determined
by Eq. (\ref{constraint on f1 (at initial time)})} [Assumption \#2b]\textit{%
. \ }If the fluid velocity $\mathbf{V(r},t)$ is bounded in the domain $%
\overline{\Omega },$ the constraint (\ref{constraint on f1 (at initial time)}%
) implies necessarily that the subdomain of velocity space in which $f_{1}$
remains strictly positive is generally a bounded subset of $%
\mathbb{R}
^{3}.$ Hence, this prescription generally corresponds to an initial 1-point
PDF which is locally \textit{non-Gaussian}, specifically because of the
(missing) tails of the PDF.

\

\subsection{3A - Solution of the initial-value problem for $f_{1}$}

\ In this case, let us consider (at $t=t_{o}$) $f_{1}$ of the general form%
\begin{equation}
f_{1}(t)=\left\langle f_{1}(t)\right\rangle _{\Omega }\frac{h(t)}{%
\left\langle h(t)\right\rangle _{\Omega }},  \label{NON-Maxwellian}
\end{equation}%
with $\left\langle f_{1}(t_{o})\right\rangle _{\Omega }$ determined by Eq.(%
\ref{constraint on f1 (at initial time)})\textit{\ }and\textit{\ }$h(t_{o})$
to be assumed again as a strictly positive and regular real function. Then
considering variations of $f_{1}$ the type%
\begin{equation}
\delta f_{1}=\left\langle f_{1}\right\rangle _{\Omega }\frac{\delta h}{%
\left\langle h\right\rangle _{\Omega }},  \label{variation}
\end{equation}%
i.e., defined so that there results identically $\delta \left\langle
f_{1}\right\rangle _{\Omega }=\delta \left\langle h\right\rangle _{\Omega
}\equiv 0,$ the variational principle (\ref{A-2}) (PEM) requires in this
case that at the initial time $t_{o},$ $h(t_{o})$ must fulfill the
variational equation
\begin{eqnarray}
&&\left. \int\limits_{\Gamma }d\mathbf{x\delta }h(t_{o})\frac{\left\langle
f_{1}\right\rangle _{\Omega }}{\left\langle h\right\rangle _{\Omega }}%
\left\{ 1+\ln h(t_{o})+\right. \right. \\
&&\left. \left. +\lambda _{o}+\mathbf{\lambda }_{1}\cdot \mathbf{u+\lambda }%
_{2}\mathbf{u}^{2}\right\} =0.\right.  \notag
\end{eqnarray}%
Thus, $f_{1}(t_{o})$ is necessarily of the form (\ref{NON-Maxwellian}),
while at $t=t_{o}$ the function $h(t)$ reads
\begin{equation}
h(t)=\exp \left\{ -1-\lambda _{o}-\mathbf{\lambda }_{1}\cdot \mathbf{u-}%
\lambda _{2}\mathbf{u}^{2}\right\} ,  \label{solution for h}
\end{equation}%
with the Lagrange multipliers $\lambda _{o},\mathbf{\lambda }_{1}$\textbf{\ }%
and\textbf{\ }$\mathbf{\lambda }_{2}$ to be determined again imposing the
moment equations (\ref{MOMENTS}). It follows that:

\begin{enumerate}
\item due to the arbitrariness of the choice of $t_{o},$ thanks to Axioms
\#3 (conservation of the BS entropy) and 4 (PEM), it follows that for all $%
t\in I,$ $f_{1}(t)$ is necessarily of the form (\ref{NON-Maxwellian});

\item unless there results identically $\frac{\left\langle
f_{1}\right\rangle _{\Omega }}{\left\langle h\right\rangle _{\Omega }}=1,$
the initial PDF $f_{1}(t_{o})$ is \textit{generally not a Gaussian, }and
hence of the form $f_{1}(t)=g_{1}(t)f_{M}(t),$with $g_{1}(t)\neq 1$ to be
assumed a suitably smooth ordinary function;

\item $f_{1}(t_{o})$ is necessarily a conditional observable (Assumption
\#1).
\end{enumerate}

\subsection{3B - The Liouville evolution equation for $f_{1}$}

To determine the time evolution of the PDF we require again that the mean
field force $\mathbf{F}(f_{1})\equiv \mathbf{F}(\mathbf{x,}t,\mathbf{\alpha }%
;f_{1})$ is a conditional observable. To determine explicitly $\mathbf{F}%
(f_{1}),$ lest us require, first, that when Eq.(\ref{Maxwellian case}) holds
identically, $\mathbf{F}(f_{1})$ must coincide with $\mathbf{F}(f_{M})$
[defined by Eqs.(\ref{F-tor})-(\ref{DEALTA f})]. The same manifestly must
occur if $f_{1}$ remains, in the whole phase space $\Gamma $, suitably
"near" to $f_{M}.$ In fact, in this case in the same set $\mathbf{F}(f_{1})$
and $\mathbf{F}(f_{M})$ must remain close too (again in some suitable
asymptotic sense). This delivers for $\mathbf{F}(f_{1})$ a prescribed
polynomial representation in terms of the relative kinetic velocity $\mathbf{%
u.}$ In principle, in fact, $\mathbf{F}(f_{1})$ might include also
polynomials of higher-degree in $\mathbf{u.}$ These terms, however, must
necessarily vanish identically for $f_{1}\equiv f_{M}$ and cannot contribute
to the moment equations [of IKE] which yield INSE. Since their form remains
therefore arbitrary, they are manifestly non-observables. \textit{Hence such
terms are ruled out by the requirement of} $\mathbf{F}(f_{1})$ \textit{being
a conditional observable}. This implies that $\mathbf{F}(f_{1})$ must be
necessarily a polynomial of second degree in the relative kinetic velocity $%
\mathbf{u,}$ whose precise form is prescribed by imposing that the moment
equations of IKE for $G(\mathbf{v,r},t)=1\mathbf{,v,\rho }_{o}u^{2}/3$ must
necessarily coincide with INSE [ Eqs.(\ref{1b})-(\ref{1bbb}) in Appendix A].
Let us introduce now the mean-field force:%
\begin{eqnarray}
\mathbf{F}(\mathbf{x,}t,\mathbf{\alpha };f_{1}) &=&\mathbf{F}_{0}(\mathbf{x,}%
t,\mathbf{\alpha };f_{1})+\mathbf{F}_{1}(\mathbf{x,}t,\mathbf{\alpha }%
;f_{1})+  \label{F} \\
&&+\Delta \mathbf{F}(\mathbf{x,}t,\mathbf{\alpha };f_{1}),
\end{eqnarray}%
$\Delta \mathbf{F}(\mathbf{x,}t,\mathbf{\alpha };f_{1})$ denoting here an
arbitrary gauge vector field of the form
\begin{equation*}
\Delta \mathbf{F}(\mathbf{x,}t,\mathbf{\alpha };f_{1})=\Delta \mathbf{F}_{1}(%
\mathbf{x,}t,\mathbf{\alpha };f_{M})+\Delta \mathbf{F}_{2}(\mathbf{x,}t,%
\mathbf{\alpha };f_{1}),
\end{equation*}%
defined so that: a) if $f_{1}\equiv f_{M},$ it coincides with
\begin{equation*}
\Delta \mathbf{F}(\mathbf{x,}t,\mathbf{\alpha };f_{1})=\Delta \mathbf{F}_{1}(%
\mathbf{x,}t,\mathbf{\alpha };f_{M});
\end{equation*}
b) it does not contribute to the moment equations of IKE, evaluated for $G(%
\mathbf{v,r},t)=1\mathbf{,v,\rho }_{o}u^{2}/3,$ namely it is defined so that
there results identically%
\begin{equation}
\int d^{3}v\Delta \mathbf{F}(\mathbf{x,}t,\mathbf{\alpha };f_{1})f_{1}=%
\mathbf{0}.
\end{equation}%
Hence $\Delta \mathbf{F}(\mathbf{x,}t,\mathbf{\alpha };f_{1})$ satisfies
manifestly the gauge condition (\ref{gauge property}). The indeterminacy of $%
\mathbf{F}(\mathbf{x,}t,\mathbf{\alpha };f_{1})$ can avoided again by
imposing again the requirement that it is a conditional observable
(Assumption \#2), namely
\begin{equation}
\Delta \mathbf{F}(\mathbf{x,}t,\mathbf{\alpha };f_{1})\equiv 0.
\label{CONSTRAINT ON F}
\end{equation}%
Here the vector fields $\mathbf{F}_{0}$ and $\mathbf{F}_{1}$ are \cite%
{Ellero2005}
\begin{eqnarray}
&&\left. \mathbf{F}_{0}\mathbf{(x,}t,\mathbf{\alpha };f_{1})=\frac{1}{\rho
_{0}}\left[ \mathbf{\nabla \cdot }\underline{\underline{{\Pi }}}-\mathbf{%
\nabla }p_{1}+\mathbf{f}_{R}\right] +\right.  \label{F-2} \\
&&+\frac{1}{2}\mathbf{u}\cdot \nabla \mathbf{V}+\frac{1}{2}\nabla \mathbf{%
V\cdot \mathbf{u}+}\nu \nabla ^{2}\mathbf{V,}  \notag
\end{eqnarray}%
\begin{eqnarray}
&&\left. \mathbf{F}_{1}\mathbf{(x,}t,\mathbf{\alpha };f_{1})=\frac{1}{2}%
\mathbf{u}\left\{ A\mathbf{(r,}t;f)\mathbf{+}\frac{1}{p_{1}}\mathbf{\nabla
\cdot Q}-\right. \right.  \label{F-3} \\
&&\left. -\left. \frac{1}{p_{1}^{2}}\left[ \mathbf{\nabla \cdot }\underline{%
\underline{\Pi }}\right] \mathbf{\cdot Q}\right\} +\frac{v_{th}^{2}}{2p_{1}}%
\mathbf{\nabla \cdot }\underline{\underline{\Pi }}\left\{ \frac{u^{2}}{%
v_{th}^{2}}-\frac{3}{2}\right\} ,\right.  \notag
\end{eqnarray}%
with $\mathbf{Q}$ and $\underline{\underline{{\Pi }}}$ denoting the moments%
\begin{equation}
\left\{
\begin{array}{c}
\mathbf{Q}\ =\rho _{o}\int d^{3}v\mathbf{u}\frac{u^{2}}{3}f_{1}, \\
\underline{\underline{{\Pi }}}=\rho _{o}\int d^{3}v\mathbf{uu}f_{1},%
\end{array}%
\right.  \label{F-4}
\end{equation}%
and the scalar field $A\mathbf{(r,}t,\mathbf{\alpha };f_{1})$ defined by Eq.(%
\ref{A field}).

\subsection{3C - Properties of $\left\{ f_{1},\Gamma \right\} $}

Let us now prove that the IKT statistical model $\left\{ f_{1},\Gamma
\right\} $ satisfies both Problems \#1 and 2 also for non-Gaussian PDF's of
the type (\ref{NON-Maxwellian}). As a consequence, the following theorem
holds:\smallskip \medskip

\textbf{THM.2 - Uniqueness and statistical completeness of }$\left\{
f_{1},\Gamma \right\} $

\emph{In validity of Assumptions \#1, 2 and 3 (i.e., 3a-3c), let us impose
that the statistical model }$\left\{ f_{1},\Gamma \right\} $ \emph{is\
defined by requiring that }$f_{1}$ \emph{is a strictly positive ordinary
function} \emph{which satisfies the initial condition (\ref{NON-Maxwellian})
with (\ref{solution for h}) and IKE (\ref{Liouvlle eq}), with the
corresponding dean field force defined by Eqs.(\ref{F})-(\ref{F-3}). Then }$%
\left\{ f_{1},\Gamma \right\} $ \emph{is unique and statistically complete. }

\emph{PROOF}

In fact, first, the initial condition (\ref{NON-Maxwellian})\emph{\ }%
determined imposing PEM together with the constraints\emph{\ }(\ref{1ca})
and (\ref{constraint on f1 (at initial time)}) is manifestly unique and
hence a $f_{1}(t_{o})$ is necessarily conditional observable. Furthermore,
thanks to Assumption \#2\emph{\ }$\mathbf{F}(\mathbf{x,}t;f_{1})$ is
necessarily of the\ form provided by Eqs. (\ref{F})-(\ref{F-3}), i.e.,
unique,\emph{\ } so that $\left\{ f_{1},\Gamma \right\} $ is unique too.
Second, the initial PDF satisfies by construction also to the\textit{\ }%
requirement posed by Eq.(\ref{constraint on f1 (at initial time)}). Let us
now prove that $f_{1}$ fulfills also the deterministic limit (\ref{Dirac
Delta}). \ This follows invoking Eq.(\ref{Dirac delta-2}) and noting that in
the limit $p_{1}\rightarrow 0^{+}$ and $\left\vert \mathbf{V}\right\vert
\rightarrow 0^{+}$\ there results manifestly%
\begin{equation}
\lim_{\substack{ p_{1}\rightarrow 0^{+}  \\ \left\vert \mathbf{V}\right\vert
\rightarrow 0^{+}}}\frac{h(t)}{\left\langle h(t)\right\rangle _{\Omega }}=1.
\end{equation}%
This proves that $\left\{ f_{1},\Gamma \right\} $ is statistically complete
too. Q.E.D.

We remark that:

\begin{itemize}
\item the proof that $\left\{ f_{1},\Gamma \right\} $ provides a complete
IKT for the INSE problem was reached previously in Ref. \cite{Ellero2005}.
This follows by noting that, by construction, (thanks to Axioms \#0-2) the
fluid fields $\left\{ \mathbf{V,}p_{1}\right\} $ necessarily satisfy the
INSE initial-boundary value problem [see Appendix A], when they are
identified with the two velocity moments of the PDF evaluated for $G=\mathbf{%
v,\rho }_{o}u^{2}/3.$

\item THM.2 \ yields a solution to Problem \#1, namely that the initial the
1-point PDF coincides with the observable defined by the initial VDFD $%
\widehat{f}_{1}^{(freq)}(\mathbf{v}_{j}\mathbf{,}t_{o},\mathbf{\alpha };Z)$
[see Eq.(\ref{constraint on f1 (at initial time)})];

\item In particular, \ THM.2 (as also THM.1) holds \textit{both in the case
in which the fluid fields are deterministic and stochastic, i.e.,} both for
the deterministic and stochastic INSE problems [defined in Appendix A];

\item In case of THM.2 the assumption that both the $1-$point PDF $f_{1}$
and $\mathbf{F}$ are conditional observables (Assumptions \#1 and 2) is
ultimately demanded also by consistency with the physical requirement posed
by the constraint (\ref{constraint on f1 (at initial time)}). In fact, it is
obvious that otherwise the uniqueness of $\left\{ f_{1},\Gamma \right\} $
might not be warranted. In particular, this means that configuration-space
average of the PDF, i.e., either the continuous or discrete averages $%
\left\langle f_{1}\right\rangle _{\Omega }$ and $\overline{f_{1}}$ might not
be unique, and hence these quantities \emph{would not be observables};

\item In Refs. \cite{Tessarotto2007a,Tessarotto2007b} the uniqueness of the
mean-field force $\mathbf{F}(\mathbf{x,}t,\mathbf{\alpha };f_{1})$\ for a
non-Gaussian $f_{1},$ was achieved again based both on phenomenological
arguments, i.e., besides the comparison with extended thermodynamics, the
requirement that $\mathbf{F}$ depends only on the minimum number of
higher-order moments of $f_{1},$ defined so that: a) they provide the
correct fluid equations; b) they vanish identically in the case (\ref%
{Maxwellian case}). THM.2 shows that the uniqueness [of $\mathbf{F}(\mathbf{%
x,}t,\mathbf{\alpha };f_{1})$] is achieved simply based on the physical
prescription that both $f_{1}$ and $\mathbf{F}(\mathbf{x,}t,\mathbf{\alpha }%
;f_{1})$ are conditional observables (see again Assumptions \#1 and 2).
\end{itemize}

\subsection{3D - Solution of the closure problem}

As indicated above, a desired property of statistical models in fluid
dynamics would be the (possible) fulfillment of suitable closure conditions,
permitting to assure that the relevant PDF advances in time by means of a
statistical equation which depends solely on the same PDF. Here we wish to
investigate the closure problem earlier posed [Subsections 1B and 1D] and,
in particular, that specified by Problems 4 and 5. It is immediate to prove
that a \textit{formal exact solution }for such closure problems is provided
by the IKT statistical model $\left\{ f_{1},\Gamma \right\} $ defined in the
previous sections 2 and 3. In other words, $\left\{ f_{1},\Gamma \right\} $
is necessarily closed, i.e., the following result holds:\smallskip \medskip

\textbf{THM.3 - Closure properties of }$\left\{ f_{1},\Gamma \right\} $ (%
\textbf{Closure Theorem})

\emph{In validity of Assumptions \#1, 2 and 3 (i.e., 3a-3c), the statistical
model }$\left\{ f_{1},\Gamma \right\} ,$\emph{\ defined by imposing the
initial condition (\ref{NON-Maxwellian}) with (\ref{solution for h}), and
with }$f_{1}(\mathbf{r,v},t)$ \emph{assumed as a strictly positive ordinary
function} \emph{satisfying IKE (\ref{Liouvlle eq}), with mean field force
defined by Eqs.(\ref{F})-(\ref{F-3}), is closed. }

\emph{PROOF}

Due to the prescription (\ref{F-tor})-(\ref{A field}) of the mean-field
force $\mathbf{F}(\mathbf{x,}t,\mathbf{\alpha };f_{1})$ it is immediate to
prove that the moment equations of IKE for $G=1,\mathbf{v},\rho _{o}u^{2}/3,$
namely%
\begin{equation}
\int\limits_{U}d^{3}\mathbf{v}GLf_{1}=0  \label{moment equations}
\end{equation}%
coincide respectively with Eq.(\ref{1ba}) (for the moments $G=1$ and $\rho
_{o}u^{2}/3)$ and with Eq.(\ref{1bbb}) (for the second moment $G=\mathbf{v}$%
), i.e., with the complete set of PDE defined by INSE [see Eqs. (\ref{1b})- %
\ref{1c}), in Appendix A]. The latter, by assumption, define a closed system
of equations, hence the moment-closure condition (Problem 4) is necessarily
satisfied. Furthermore, to prove that also the\textit{\ }kinetic-closure
condition (posed by Problem 5) holds it is sufficient to notice that the
same mean-field force $\mathbf{F}(\mathbf{x,}t,\mathbf{\alpha };f_{1})$
depends, by construction, only on the fluid fields $\mathbf{V},p_{1},$ $%
\mathbf{Q},$and $\underline{\underline{\mathbf{\Pi }}}.$ Therefore, also the
requirement of kinetic-closure is necessarily satisfied. Q.E.D.

\section{4 - Comparisons with previous approaches}

Interesting issues are posed by the comparison with previous statistical
treatments and in particular the statistical model $\left\{ f_{H},\Gamma
\right\} ,$ underlying both the HRE \cite{Hopf1952, Rosen1960,Edwards1964}
and ML \cite{Monin1967,Lundgren1967} approaches$.$This is relevant in
special reference to:

\begin{itemize}
\item analyze basic properties of $\left\{ f_{H},\Gamma \right\} $ (see
Subsection 4A);

\item determine the explicit relationship between the PDF's $f_{1}$ and $%
f_{H},$ which characterize the two approaches (Subsection 4B);

\item the construction of the statistical equation for the
stochastic-average of $f_{1}$ (Subsection 4C),

\item the comparison with the HRE functional-differential approach
(Subsection 4D).
\end{itemize}

\subsection{4A - Properties of $\left\{ f_{H},\Gamma \right\} $}

It is immediate to show that $\left\{ f_{H},\Gamma \right\} :$ \ 1) holds
both for the deterministic and stochastic INSE problems [see Appendix A]; 2)
realizes an IKT for INSE, which is \textit{unique} and \textit{closed; }3)
it is (generally) not a\textit{\ complete IKT }(in fact, manifestly, neither
the fluid pressure, nor the kinetic pressure, can be represented in terms of
velocity moments of $f_{H}$); 4) in addition, it \textit{is }(generally)%
\textit{\ not statistically complete}. Indeed, since $f_{H}$ is a
distribution, its configuration-space average cannot generally expected to
agree with the the observable $\widehat{f}_{1}^{(freq)}(\mathbf{v,}t,\mathbf{%
\alpha };Z)$ or VDFD ($1-$point velocity-frequency density function). The
following result holds:\smallskip \medskip

\textbf{THM.4 - Statistical incompleteness of}\textit{\ }$\left\{
f_{H},\Gamma \right\} $

\emph{The statistical model }$\left\{ f_{H},\Gamma \right\} $\emph{\ does
not generally fulfill the constraint (\ref{constraint on f1 (at initial
time)}).}

PROOF

To reach the proof, \ let us evaluate the configuration-space average of $%
f_{H}.$ For definiteness, let us adopt the definition of discrete average
provided by Eq.(\ref{f_1 discrete- approximate}). Thus, by denoting $%
Z_{i}\equiv \left\{ \mathbf{V}_{i}(t),p_{i}(t)\right\} $ the average fluid
fields in the $i-$th cell of $\Omega $ (and evaluated at position $\mathbf{r}%
_{i})$, it follows that the configuration-space average of $f_{H}$ reads
\begin{equation}
\left\langle f_{H}(t)\right\rangle _{\Omega }=\frac{1}{N_{\ast }}%
\sum\limits_{i=1,N_{\ast }}\delta (\mathbf{v-V}_{i}(t)),
\end{equation}%
i.e., $\left\langle f_{H}\right\rangle _{\Omega }(\mathbf{v},t)$ \textit{is
still a distribution. Hence, it cannot generally be identified with the
observable }$\widehat{f}_{1}^{(freq)}$ (which by assumption here it is
considered as an ordinary function)! For comparison, instead, the complete
IKT approach yields instead:%
\begin{equation}
\left\langle f_{1}(t)\right\rangle _{\Omega }=\frac{1}{N_{\ast }}%
\sum\limits_{i=1,N_{\ast }}f_{1}(t).
\end{equation}%
where in this case by definition $f_{1}$ is an ordinary function (and hence
its configuration-space average $\left\langle f_{1}(t)\right\rangle _{\Omega
}$ can be identified with the observable/conditional observable $\widehat{f}%
_{1}^{(freq)}(t)$, which - on the contrary is generally an ordinary
function. Q.E.D.

A similar proof can be achieved also adopting the continuous operator (\ref%
{f_1-average}). We remark here that:

\begin{enumerate}
\item the setting based on the definition (\ref{f_1 discrete- approximate})
is actually consistent with the physical measurement process of the
corresponding 1-point velocity frequency [$\widehat{f}_{1}^{(freq)}(t)$],
based of the discretization of the fluid domain $\Omega $];

\item the extension of THM.4 to a generic stochastic model of the type $%
\left\{ \left\langle f_{H}\right\rangle ,\Gamma \right\} $ is not possible,
as proven by the subsequent discussion [see in particular THM.5].
\end{enumerate}

\subsection{4B - Relationship between $\left\{ f_{1},\Gamma \right\} $ and $%
\left\{ f_{H},\Gamma \right\} $}

Let us now pose the problem of determining the relationship between the
1-point PDF which characterizes the IKT statistical model $\left\{
f_{1},\Gamma \right\} $ and the PDF $f_{H}(t)$ associated to the statistical
model $\left\{ f_{H},\Gamma \right\} .$ \textit{We intend to show that } $%
f_{1}$ [with $f_{1}\equiv f_{1}(\mathbf{r},\mathbf{u,}t),$ $\mathbf{u=v-V}(%
\mathbf{r},t)$ and $\mathbf{x=}\left( \mathbf{r,v}\right) \in \Gamma $]
\textit{can be identified with a suitable} \textit{stochastic-average of} $%
f_{H},$ to be defined in terms of an appropriate nonhomogeneous and
non-stationary stochastic model (see related definitions in Appendix A).

To assess properly precisely the statement let us introduce for the fluid
fields $\left\{ Z\right\} $ the stochastic representation%
\begin{equation}
\left\{ Z(\Delta \mathbf{V})\right\} =\left\{ \mathbf{V}(\mathbf{r}%
,t)+\Delta \mathbf{V},p(\mathbf{r},t)\right\} ,  \label{stochastic fields}
\end{equation}%
where $\left\{ Z\right\} =\left\{ \mathbf{V}(\mathbf{r},t),p(\mathbf{r}%
,t)\right\} $ and $\Delta \mathbf{V\in
\mathbb{R}
}^{3}$ denote respectively an arbitrary particular solution of the INSE
problem [see Appendix A] and an arbitrary velocity fluctuation. Here by
assumption, the vector $\Delta \mathbf{V}\equiv (\Delta V_{1},\Delta
V_{2},\Delta V_{3})$ will be considered as a set of stochastic hidden
variables characterized by the stochastic PDF
\begin{equation}
g_{1}(\Delta \mathbf{V,r,}t)\equiv f_{1}(\mathbf{r},\Delta \mathbf{V,}t)%
\mathbf{,}  \label{definition of the stochastic PDF}
\end{equation}%
with $f_{1}$ denoting the 1-point PDF characterizing the IKT statistical
model $\left\{ f_{1},\Gamma \right\} $ previously introduced [and obeying
Eq.(\ref{Liouvlle eq}) when replacing $\Delta \mathbf{V\rightarrow v}$]$%
\mathbf{,}$ The set $\left\{ \Delta \mathbf{V,}g_{1}\right\} $ defines
therefore (a generally non-homogeneous and non-stationary) stochastic model
defined so that, by assumption, its moments are necessarily defined so that%
\begin{eqnarray}
\left\langle 1\right\rangle _{\mathbf{\alpha }} &\equiv &\int\limits_{%
\mathbb{R}
^{3}}d^{3}\Delta \mathbf{V}f_{1}(\mathbf{r},\Delta \mathbf{V,}t),
\label{ASS-1} \\
\mathbf{0} &\equiv &\int\limits_{%
\mathbb{R}
^{3}}d^{3}\Delta \mathbf{V}\Delta \mathbf{V}f_{1}(\mathbf{r},\Delta \mathbf{%
V,}t),  \label{ASS-2} \\
\frac{1}{2}\rho _{o}v_{th}^{2}(\mathbf{r},t) &\equiv &\frac{2}{3}\rho
_{o}\int\limits_{%
\mathbb{R}
^{3}}d^{3}\Delta \mathbf{V}\frac{1}{2}\left( \Delta \mathbf{V}\right)
^{2}f_{1}(\mathbf{r},\Delta \mathbf{V,}t).  \label{ASS-3}
\end{eqnarray}%
Here the moment on the r.h.s. of Eq.(\ref{ASS-3}) represents, up to the
constant factor $\frac{2}{3}\rho _{o}$, the stochastic mean value of the
\textit{stochastic kinetic energy per unit mass} $\frac{1}{2}\left( \Delta
\mathbf{V}\right) ^{2},$ while $\frac{1}{2}\rho _{o}v_{th}^{2}(\mathbf{r},t)$
and $v_{th}(\mathbf{r},t)$ are the \textit{thermal energy} and the
corresponding \textit{thermal\ velocity produced by the kinetic pressure} $%
p_{1}(\mathbf{r},t)$ [with $p_{1}(\mathbf{r},t)$ defined according to Eq. (%
\ref{kinetic pressure})]%
\begin{equation}
v_{th}(\mathbf{r},t)\mathbf{=}\sqrt{\frac{\mathbf{2}p_{1}(\mathbf{r},t)}{%
\rho _{o}}.}  \label{thermal velocoty}
\end{equation}%
It follows that:

\textbf{THM.5 - Representation of }$f_{1}$\textbf{\ in terms of} $f_{H}$

\emph{The stochastic representation (\ref{stochastic fields}) and position (%
\ref{definition of the stochastic PDF}) can always be introduced. It follows
that the stochastic average of }$f_{H},$\emph{\ defined with respect to}
\emph{the stochastic averaging operator (\ref{stochastic averaging operator})%
}$,$ \emph{reads identically}
\begin{equation}
\left\langle f_{H}\right\rangle _{\Delta \mathbf{V}}=f_{1}(\mathbf{r},%
\mathbf{u,}t)\mathbf{,}  \label{RELATIONSHIP}
\end{equation}%
\emph{with }$f_{H}$ \emph{given by Eq.(\ref{DIRAC DELTA- HRE}).}

PROOF

First we notice that if $\left\{ Z\right\} $ is an arbitrary particular
solution of INSE, also $\left\{ Z(\Delta \mathbf{V})\right\} $ is manifestly
a particular solution of the same equation. The proof is reached by
introducing in the NS equation (\ref{1bbb}) [see Appendix A] a suitable
stochastic volume force $\mathbf{f+\Delta f,}$ with $\mathbf{\Delta f}$
defined so that%
\begin{equation}
\frac{1}{\rho _{o}}\Delta \mathbf{f=\Delta V}\cdot \nabla \mathbf{V.}
\end{equation}%
Hence, the stochastic representation (\ref{stochastic fields}) always holds,
with $\Delta \mathbf{V}$ to be considered as stochastic hidden variables. In
addition, since the definition of $g_{1}(\Delta \mathbf{V,r,}t)$ is
arbitrary, the position (\ref{definition of the stochastic PDF}) can always
be introduced. These definitions uniquely prescribe the stochastic model $%
\left\{ \Delta \mathbf{V,}g_{1}\right\} $ and the related stochastic
averaging operator (\ref{stochastic averaging operator}) [see Appendix A].
It follows that $\left\langle f_{H}\right\rangle _{\Delta \mathbf{V}}\equiv
\int\limits_{%
\mathbb{R}
^{3}}d^{3}\mathbf{\Delta V}f_{1}(\mathbf{r},\Delta \mathbf{V,}t)\delta
\mathbf{(v-V}(\mathbf{\mathbf{r,}}t)-\mathbf{\Delta V})$ yields Eq.(\ref%
{RELATIONSHIP}), which is therefore identically satisfied. \ Q.E.D.

In conclusion:

\begin{itemize}
\item the 1-point PDF of the IKT statistical model can be considered simply
as a \textit{possible stochastic realization} of the PDF $f_{H},$ achieved
by means of the stochastic model $\left\{ \Delta \mathbf{V,}g_{1}\right\} ,$
in which $g_{1}$ is properly related to the 1-point PDF characterizing the
IKT statistical model;

\item in view of the positions (\ref{ASS-3}) and (\ref{thermal velocoty}), $%
\left\{ \Delta \mathbf{V,}g_{1}\right\} $ can be viewed as the \textit{%
stochastic model which takes into account the thermal motion produced in a
NS fluid by its kinetic pressure }$p_{1}(\mathbf{r},t)$ [see Eq. (\ref%
{kinetic pressure})].
\end{itemize}

\subsection{4C - Consequences - Statistical equation for $\left\langle
f_{1}\right\rangle $}

The same conclusion (i.e., THM.5) holds manifestly also in the case in which
the fluid fields $\left\{ Z\right\} $, as well as the same 1-point PDF $%
f_{1} $ and the mean-field force $\mathbf{F}(f_{1}),$ are stochastic in the
sense of Eq.(\ref{stochastic functions}), i.e., they depend on a suitable
set of stochastic hidden variables $\mathbf{\alpha \in V}_{\mathbf{\alpha }%
}\subseteq
\mathbb{R}
^{n}\mathbf{,}$ here considered independent of $\Delta \mathbf{V}$ and
characterized by a stochastic PDF $g$. For definiteness, let us assume that $%
g$ is homogeneous and stationary, i.e., of the form $g$ =$g(\mathbf{\alpha }%
).$ Then the ensemble average of the PDF, $\left\langle f_{1}\right\rangle ,$
can be identified with
\begin{equation}
\left\langle f_{1}\right\rangle \equiv \left\langle f_{1}\right\rangle _{%
\mathbf{\alpha }},  \label{stoch-average}
\end{equation}%
with $f_{1}$ defined by Eq.(\ref{RELATIONSHIP}) and the brackets $%
\left\langle \cdot \right\rangle _{\mathbf{\alpha }}$ denoting the
stochastic averaging operator (\ref{stochastic averaging operator})
[Appendix A]. Hence, as a consequence of THM.5 and IKE, it follows that $%
\left\langle f_{1}\right\rangle $ obeys necessarily \textit{the
stochastic-averaged IKE} \cite{Tessarotto2008-4,Tessarotto2009}%
\begin{eqnarray}
&&\left. \frac{\partial }{\partial t}\left\langle f_{1}\right\rangle +%
\mathbf{v\cdot }\frac{\partial }{\partial \mathbf{r}}\left\langle
f_{1}\right\rangle +\frac{\partial }{\partial \mathbf{v}}\cdot \left\{
\left\langle \mathbf{F}(f_{1})\right\rangle \left\langle f_{1}\right\rangle
\right\} =\right.  \label{stochastic-averaged Liouville equation} \\
&=&-\frac{\partial }{\partial \mathbf{v}}\cdot \left\{ \left\langle \delta
\mathbf{F}(f_{1})\delta f_{1}\right\rangle \right\} .
\end{eqnarray}%
Here $\delta f_{1}$ and $\delta \mathbf{F}(f_{1})$ denote the stochastic
fluctuations [see Eqs.(\ref{stochastic decomposition 2}) and \ (\ref%
{stochastic decomposition 3}) in Appendix A], while $\mathbf{F}(f_{1})$ is
defined by Eqs. (\ref{F}),(\ref{CONSTRAINT ON F}),(\ref{F-2})-(\ref{F-4})
and (\ref{A field}). \ We remark that Eq.(\ref{stochastic-averaged Liouville
equation}) takes into account the constraints placed on $\left\langle
f_{1}\right\rangle $ by the physical realizability conditions (see Section
2), as well the axioms of IKT (Section 2). Hence:

\begin{itemize}
\item it is appropriate for describing homogenous and stationary turbulence
in NS fluids;

\item however, it can be generalized, in principle, to a generally
non-homogeneous and non-stationary stochastic PDF $g,$ as appropriate for
describing non-homogeneous and non-stationary turbulence \cite%
{Tessarotto2009}.
\end{itemize}

The equation \textit{departs from the statistical }(or \textquotedblright
transport\textquotedblright\ )\textit{\ equation} [for $\left\langle
f_{1}\right\rangle $] \textit{usually considered in the literature} [see,
for example, Dopazo \cite{Dopazo} and Pope \cite{Pope2000}]. This fact is
not surprising. In fact, unlike the customary approaches, here:

\begin{itemize}
\item the operator of ensemble average "$\left\langle {}\right\rangle ",$
defined by Eqs. (\ref{RELATIONSHIP}) and (\ref{stoch-average}), does not
commute with the relevant differential operators ($\frac{\partial }{\partial
t},\nabla ,\nabla ^{2});$

\item by assumption, the 1-point PDF satisfies the Liouville statistical
equation defined by (\ref{Liouvlle eq}), together with the physical
realizability conditions imposed on the 1-point PDF [see Eqs. (\ref{MOMENTS}%
)-(\ref{kinetic pressure}),(\ref{constraint on f1 (at initial time)}) and (%
\ref{Dirac Delta})]. These constraints are not required in the customary
approach \cite{Dopazo,Pope2000};

\item the definitions of the vector field $\mathbf{F}(f_{1})$ here adopted
satisfy the requirements placed by the axiomatic formulation (see Section 2,
Subsection 2A).
\end{itemize}

\subsection{4D - Relationship between $\left\{ f_{1},\Gamma \right\} $ and
the HRE functional-differential approach}

Finally let us consider the comparison with the HRE functional-differential
approach \cite{Hopf1952,Rosen1960,Edwards1964}. In analogy to Ref.\cite%
{Hosokawa2006} let us introduce the functional%
\begin{equation}
\phi \left[ y(x),t\right] =\int\limits_{\Gamma }d\mathbf{x}y(\mathbf{x}%
)f_{1}(\mathbf{x},t),
\end{equation}%
which implies
\begin{eqnarray}
\frac{\partial }{\partial t}\phi \left[ y(x),t\right] &=&-\int\limits_{%
\Gamma }d\mathbf{x}y(\mathbf{x})\left[ \mathbf{v\cdot }\frac{\partial }{%
\partial \mathbf{r}}f_{1}(\mathbf{x},t)+\right. \\
&&\left. \frac{\partial }{\partial \mathbf{v}}\cdot \left\{ \mathbf{F}(%
\mathbf{x},t;f_{1})f_{1}(\mathbf{x},t)\right\} \right] .
\end{eqnarray}%
Then it follows that $\phi \left[ y(x),t\right] $ must obey the following
single linear functional-differential equation, manifestly equivalent to IKE
(Liouville equation) (\ref{Liouville op}),%
\begin{equation}
\frac{\partial }{\partial t}\phi \left[ y(x),t\right] =-\int\limits_{\Gamma
}d\mathbf{x}y(\mathbf{x})Q\frac{\delta \phi }{\delta y(\mathbf{x})},
\label{Fi-equation}
\end{equation}%
where $Q$ is related to the Liouville operator and is defined as
\begin{equation}
Q\mathbf{\cdot }=\mathbf{v\cdot }\frac{\partial }{\partial \mathbf{r}}%
\mathbf{\cdot }+\frac{\partial }{\partial \mathbf{v}}\cdot \left\{ \mathbf{F}%
(\mathbf{x},t;f_{1})\mathbf{\cdot }\right\} .
\end{equation}%
Eq.(\ref{Fi-equation}) is analogous to the Hopf $\phi -$equation and has the
\textit{explicit exact solution}%
\begin{equation}
\phi \left[ y(x),t\right] =\phi \left[ T_{t_{o},t}y(x_{o}),t_{o}\right] ,
\end{equation}%
where $T_{t_{o},t}$ is the evolution operator associated to the
Navier-Stokes dynamical system.

\section{5 - IKT for multi-point PDF's}

The construction of multi-point PDF's is a problem of "practical" interest
in experimental/numerical research in fluid dynamics. In fact, despite not
being themselves observables, they are nevertheless related to physical
observables (or conditional observables), such as the velocity difference
between different fluid elements usually adopted for the statistical
analysis of turbulent fluids.

In the present theory, unlike the ML approach, the statistical equation
advancing in time the $1-$point PDF $f_{1}$ [i.e., Eq.(\ref{Liouvlle eq})]
satisfies, by definition, a kinetic closure condition. Hence, the
construction of the multi-point PDF's is trivial. Nonetheless, it is still
useful to analyze elementary implications (of the present theory) dealing
with: a) the specific representation of certain "reduced" multi-point PDF's,
defined in terms of the $1-$point PDF; b) their dynamics, namely the
statistical equations which they fulfill; c) their relationship with the
relevant observables.

\subsection{5A - Liouville equations for the multi-point PDF's}

Let us assume, for definiteness, that $f_{1}(\mathbf{x}_{i}\mathbf{,}t,%
\mathbf{\alpha };Z)$ is the $1-$point PDF which is particular solution of
the ($1-$point) Liouville equation (\ref{Liouvlle eq}). Then, denoting $%
f_{1}(i)\equiv f_{1}(\mathbf{x}_{i}\mathbf{,}t,\mathbf{\alpha };Z)$ (for $%
i=1,s$) the same PDF evaluated at the states $\mathbf{x}_{i}\equiv \left(
\mathbf{r}_{i},\mathbf{v}_{i}\right) $ (for $i=1,s$)$,$ the $s-$point PDF is
the probability density
\begin{equation}
f_{s}(1,2,..s)\equiv \prod\limits_{i=1,s}f_{1}(i),
\end{equation}%
defined in the product phase-space $\Gamma ^{s}\equiv
\prod\limits_{i=1,s}\Gamma .$ The statistical equation advancing in time $%
f_{s}$ follows trivially from Eq.(\ref{Liouvlle eq}). In fact, denoting by $%
\mathbf{F}(i)\equiv \mathbf{F}(\mathbf{x}_{i},t;f_{1})$ the mean-field force
at the state $\mathbf{x}_{i}$ (for $i=1,s$) and introducing the $s-$point
Liouville operator (with summation understood on repeated indexes)
\begin{equation}
L_{s}(1,..,s)\equiv \frac{\partial }{\partial t}+\mathbf{v}_{i}\mathbf{\cdot
}\frac{\partial }{\partial \mathbf{r}_{i}}+\frac{\partial }{\partial \mathbf{%
r}_{i}}\cdot \left\{ \mathbf{F}_{i}(i)\right\} ,
\label{s-point Liouville op}
\end{equation}%
it follows that $f_{s}(1,2,..s)$ satisfies identically the $s-$point\textit{%
\ Liouville equation} (or IKE)%
\begin{equation}
L_{s}(1,..,s)f_{s}(1,2,..s)=0.  \label{s-point Liouville equation}
\end{equation}

\subsection{5B - Consequences: reduced 2-point PFD's and 2-point observables}

In terms of the 2-point PDF, $f_{2}(1,2)$, a number of reduced probability
densities can be defined in suitable subspaces of $\Gamma ^{2}$. To
introduce them explicitly let us first introduce the transformation to the
center of mass coordinates of the two point-particles with states $\left(
\mathbf{r}_{i},\mathbf{v}_{i}\right) $ (for i$=1,2$)
\begin{equation}
\left\{ \mathbf{r}_{1},\mathbf{v}_{1},\mathbf{r}_{2},\mathbf{v}_{2}\right\}
\rightarrow \left\{ \mathbf{r,R,v,V}\right\}
\end{equation}%
[with $\mathbf{r=\frac{\mathbf{r}_{1}-\mathbf{r}_{2}}{2},R=}\frac{\mathbf{r}%
_{1}+\mathbf{r}_{2}}{2},\mathbf{v=\frac{\mathbf{v}_{1}-\mathbf{v}_{2}}{2}}$
and $\mathbf{V=}\frac{\mathbf{v}_{1}+\mathbf{v}_{2}}{2}$]. Then, these are
respectively the\textit{\ local }(in configuration space)\textit{\
velocity-difference 2-point PDF}:%
\begin{equation}
\left.
\begin{array}{c}
g_{2}(\mathbf{r}_{1},\mathbf{r}_{2},\mathbf{v},t,\mathbf{\alpha }%
)=\int_{U}d^{3}\mathbf{V}f_{2}(1,2)\equiv \\
\equiv \int d^{3}\mathbf{V}f_{1}(\mathbf{r}_{1}\mathbf{,v+V,}t,\mathbf{%
\alpha }))f_{1}(\mathbf{r}_{2},\mathbf{V-v,}t,\mathbf{\alpha };Z)%
\end{array}%
\right.  \label{g2}
\end{equation}%
(defined in the space $\Omega ^{2}\times U$) and the \textit{%
velocity-difference 2-point PDF, }i.e., the observable
\begin{equation}
\widehat{f}_{2}(\mathbf{r},\mathbf{v,}t,\mathbf{\alpha })=\left\langle g_{2}(%
\mathbf{r+R},\mathbf{r-R},\mathbf{v},t,\mathbf{\alpha })\right\rangle _{%
\mathbf{R,}\Omega }  \label{2-point velocity difference PDF}
\end{equation}%
(defined in $\Gamma $), $\left\langle {}\right\rangle _{\mathbf{R},\Omega }$
denoting the configuration-space average operator acting on the center of
mass. Hence, in terms of the average operator (\ref{f_1-average}), there
follows
\begin{eqnarray}
&&\left. \widehat{f}_{2}(\mathbf{r},\mathbf{v,}t,\mathbf{\alpha })=\frac{1}{%
\mu (\Omega )}\int_{\Omega }d^{3}\mathbf{R}\right. \\
&&g_{2}(\mathbf{r+R},\mathbf{r-R},\mathbf{v},t,\mathbf{\alpha }).  \notag
\end{eqnarray}%
In the case of a Gaussian PDF [(\ref{Maxwellian case})], Eq.(\ref{g2})
delivers in particular the Gaussian PDF%
\begin{equation}
g_{2}(\mathbf{r}_{1},\mathbf{r}_{2},\mathbf{v},t,\mathbf{\alpha })=\frac{1}{%
\pi ^{3/2}v_{th}^{3}}\exp \left\{ -\frac{\left\Vert \mathbf{v-}\frac{\mathbf{%
V}(1)-\mathbf{V}(2)}{2}\right\Vert ^{2}}{v_{th}^{2}}\right\}
\label{GAUSSIAN 2-POINT PDF}
\end{equation}%
where $\mathbf{V}(i)\equiv \mathbf{V}(\mathbf{r}_{i},t),$ $%
v_{th,p}^{2}(i)=v_{th,p}^{2}(\mathbf{r}_{i},t)$ and $v_{th}^{2}$ denotes

\begin{equation}
v_{th}^{2}=\frac{v_{th,p}^{2}(1)+v_{th,p}^{2}(2)}{4}.
\end{equation}

In a similar way it is possible to identify additional 2-point observables.
Precisely these can be defined as: \textit{\ }

\begin{enumerate}
\item \textit{the velocity-difference 2-point PDF for parallel velocity
increments}$.$\textit{\ }Introducing the representations $\mathbf{v}=\mathbf{%
n}v$ and $\mathbf{r}=\mathbf{n}r,$ $\mathbf{n}$ denoting a unit vector, $%
\widehat{f}_{2\parallel }(r,v\mathbf{,}t)$ can be simply defined as the
solid-angle average%
\begin{equation}
\widehat{f}_{2\parallel }(r,v\mathbf{,}t,\mathbf{\alpha })=\int d\Omega (%
\mathbf{n})\widehat{f}_{2}(\mathbf{r=n}r,\mathbf{v}=\mathbf{\mathbf{n}}v%
\mathbf{,}t,\mathbf{\alpha });
\label{2-point velocity-difference PDF for parallel}
\end{equation}

\item \textit{the velocity-difference 2-point PDF for perpendicular velocity
increments}$.$\textit{\ } Introducing, instead, the representations $\mathbf{%
v}=\mathbf{n}v$ and $\mathbf{r}=\mathbf{n\times b}r,$ $\mathbf{n}$ and $%
\mathbf{b}$ denoting two independent unit vectors, $\widehat{f}_{2\perp }(r,v%
\mathbf{,}t)$ can be defined as the double-solid-angle average%
\begin{eqnarray}
&&\left. \widehat{f}_{2\perp }(r,v\mathbf{,}t,\mathbf{\alpha })=\int d\Omega
(\mathbf{n})\int d\Omega (\mathbf{b})\right.
\label{2-point velocity-difference PDF for perpendicular} \\
&&\left. \widehat{f}_{2}(\mathbf{r=n\times b}r,\mathbf{v}=\mathbf{\mathbf{n}}%
v\mathbf{,}t,\mathbf{\alpha }).\right.  \notag
\end{eqnarray}

An interesting property which emerges from these definitions is that in all
cases indicated above [i.e., Eqs.(\ref{2-point velocity difference PDF}),(%
\ref{2-point velocity-difference PDF for parallel}) and (\ref{2-point
velocity-difference PDF for perpendicular})] the definition of $g_{2}$ given
above [Eq.(\ref{g2})] implies that non-Gaussian features, respectively in $%
\widehat{f}_{2},\widehat{f}_{2\parallel }$ and $\widehat{f}_{2\perp },$ may
arise even if the $1-$point PDF is Gaussian, i.e., the requirement (\ref%
{Maxwellian case}) holds identically. This occurs due to velocity and
pressure fluctuations occurring between different spatial positions $\mathbf{%
r}_{1}$ and $\mathbf{r}_{2}.$More generally. however, we can infer that -
due to the constraint (\ref{constraint on f1 (at initial time)}) here
imposed on the 1-point PDF - it is obvious that, if the fluid velocity $%
\mathbf{V}(\mathbf{r},t)$ is bounded in the domain $\overline{\Omega },$%
\textit{\ the same 1-point PDF, and hence the 2-point PDF's, cannot be
Gaussian distributions}.
\end{enumerate}

\subsection{5C - Statistical evolution equation for $\widehat{f}_{2}$}

From the $2-$point IKE (\ref{s-point Liouville equation}) (obtained in the
case $s=2$) it is immediate to obtain the corresponding evolution equation
for the reduced PDS's indicated above. For example, the velocity-difference
2-point PDF $\widehat{f}_{2}$ satisfies the equation
\begin{eqnarray}
&&\frac{\partial \widehat{f}_{2}}{\partial t}+\mathbf{v\cdot }\frac{\partial
}{\partial \mathbf{r}}\widehat{f}_{2}+\frac{\partial }{\partial \mathbf{v}}%
\cdot \frac{1}{\mu (\Omega )}\int d^{3}\mathbf{V}  \label{F-P-1} \\
&&\left. \int_{\Omega }d^{3}\mathbf{R}\frac{\mathbf{F}_{1}(1)-\mathbf{F}%
_{2}(2)}{2}f_{2}(1,2)=0.\right.  \notag
\end{eqnarray}%
It follows that, in particular, in the Gaussian case (\ref{Maxwellian case})
this equation reduces to the (generally non-Markovian) Fokker-Planck equation%
\begin{eqnarray}
&&\frac{\partial \widehat{f}_{2}}{\partial t}+\mathbf{v\cdot }\frac{\partial
}{\partial \mathbf{r}}\widehat{f}_{2}+\frac{\partial }{\partial \mathbf{v}}%
\cdot \frac{1}{\mu (\Omega )}\int_{\Omega }d^{3}\mathbf{R}  \label{F-P-2} \\
&&\left. \mathbf{F}^{(T)}g_{2}(\mathbf{r+R},\mathbf{r-R},\mathbf{v},t,%
\mathbf{\alpha })=0,\right.  \notag
\end{eqnarray}%
where the vector field $\mathbf{F}_{1}^{(T)}\equiv \mathbf{F}_{1}^{(T)}(%
\mathbf{r}_{1}\mathbf{,r}_{2},\mathbf{V},t,\mathbf{\alpha ;}f_{M})$ is given
in Appendix C [see Eqs.(\ref{F0-B-}) and \ref{F1-B-})].

An interesting issue is provided by the comparison with the statistical
formulation developed by Peinke and coworkers \cite%
{Naert1997,Friedrich1999,Luck1999,Renner2001,Renner2002}. Their approach,
based on the statistical analysis of experimental observations, indicates
that in case of stationary and homogeneous turbulence both the 2-point PDF's
for parallel and velocity increments obey stationary Fokker-Planck
equations. In particular, according to experimental evidence \cite%
{Renner2001,Renner2002} a reasonable agreement with a Markovian
approximation for Eq.(\ref{F-P-2}) - at least in some limited subset of
parameter space- is suggested. Our theory suggests, however, that a
breakdown of the Markovian assumption should be expected due to non-local
contributions appearing in the PDF's and in the corresponding statistical
equations.

\section{6 - Discussion and concluding remarks}

An axiomatic approach, based on the IKT statistical model $\left\{
f_{1},\Gamma \right\} $, has been developed for the statistics\ of the
1-point PDF $f_{1}$ which characterizes an incompressible NS fluid. The
paper contains several new aspects and basic consequences of interest in
fluid dynamics and turbulence theory.

Indeed, the theory here developed applies both to regular and turbulent
flows, characterized respectively by deterministic and stochastic fluid
fields. In fact, in both cases the time evolution of $f_{1}$ is a Liouville
equation (IKE) [see Eq.(\ref{Liouvlle eq})] which evolves in time also the
complete set of fluid fields (represented in terms of moments of the same
PDF).

In this paper an explicit solution of the problems 1-6 posed in Subsection
1D \ has been proposed.

In particular, we have proven that - extending the statistical approach
earlier developed \cite{Tessarotto2004,Ellero2005} - the IKT statistical
model $\left\{ f_{1},\Gamma \right\} $ can be uniquely determined. The
present theory is based on two new hypotheses, i.e., A) that both $f_{1}$
and $\mathbf{F}(f_{1})$ are conditional observables and B) that $f_{1}$
satisfies suitable physical realizability conditions (see Subsection 2A).
The first requirement permits to determine the mean-field force $\mathbf{F}%
(f_{1})$, while the second one uniquely prescribes - by means of PEM (i.e.,
Axiom \#4) - the initial PDF $f_{1}(t_{o})$. \ In detail, we have shown that
$\left\{ f_{1},\Gamma \right\} $ can be constructed in such a way to be:

\begin{itemize}
\item unique (see THM.1, related to Problems 1);

\item statistically complete (see THM.2, Problem 2);

\item closed, i.e., it both moment-closure and kinetic-closure conditions
(see THM.3, Problems 3 and 4)
\end{itemize}

and furthermore:

\begin{itemize}
\item that it determines uniquely all multi-point PDF's, as well as the
related observables (see Problem 5);

\item that\textit{\ }the statistical equations for multi-point PDF's depend
only on $f_{1}$ and therefore, by definition, satisfy a closure condition
(Problem 6).
\end{itemize}

The theory has important consequences.

First, it implies that the initial PDF is generally non-Gaussian (see again
THM. 2). This conclusion holds even in the case in which the fluid fields
are deterministic, namely for regular flows. In fact, the Gaussian 1-point
PDF although unique (THM.1) does not generally provide a statistical
complete model $\left\{ f_{M},\Gamma \right\} $ (see corollary of THM.1). In
addition:

\begin{itemize}
\item thanks to the fluid and kinetic closure conditions imposed on the
statistical equation for the 1-point PDF , i.e., IKE [Eq.(\ref{Liouvlle eq}%
)], $f_{1}$ depends only on a finite set of moments of the same PDF and its
time evolution is independent of higher-order (multi-point) PDF's;

\item as a basic consequence, the \textit{exact statistical equation} for
the ensemble-averaged (or stochastic-averaged) PDF $\left\langle
f_{1}\right\rangle $ has been obtained. This is found to be intrinsically
different from the analogous transport equation obtained in the past in the
case of stationary and homogeneous turbulence \cite{Dopazo,Pope2000}.
\end{itemize}

The connection of the present theory both with previous IKT approaches \cite%
{Tessarotto2004,Ellero2005} and the HRE (Hopf, \ Rosen and Edwards \cite%
{Hopf1952,Rosen1960,Edwards1964}) and ML (Monin and Lundgren \cite%
{Monin1975,Lundgren1967}) statistical treatments,\ has been pointed out
(see, in particular, Section 4).\ Regarding, in particular, the last two
approaches the following results have been reached:

\begin{itemize}
\item the common statistical model, $\left\{ f_{H},\Gamma \right\} ,$ used
in both approaches (HRE and ML) has been shown to be generally statistically
incomplete (THM.4) $;$

\item the relationship between the $f_{1}$ and the PDF $f_{H}$ which
characterizes the HRE and ML approaches has been determined. In particular,
we have proven that $f_{1}$ can be identified with a suitable stochastic
average of $f_{H},$ via a generally non-homogeneous and non-stationary
stochastic PDF [see Eq.(\ref{RELATIONSHIP})];

\item the unique connection [via Eqs.(\ref{RELATIONSHIP}) and (\ref%
{stoch-average})] existing between the ensemble-averaged PDF's $\left\langle
f_{1}\right\rangle $ and $\left\langle f_{H}\right\rangle $ has been
displayed$;$

\item the relationship with the Hopf's functional-differential approach has
been pointed out.
\end{itemize}

Finally, as an application, explicit representations have been given for the
\textit{reduced 2-point PDF's} usually adopted for the statistical
description of turbulent flows, represented respectively by the
velocity-difference 2-point PDF $\widehat{f}_{2}$ and the
velocity-difference 2-point PDF for parallel and perpendicular velocity
increments $\widehat{f}_{2\parallel }$ and $\widehat{f}_{2\perp }.$

\section{Acknowledgements}

Work developed in the framework of the MIUR (Italian Ministry of University
and Research) PRIN Research Program \textquotedblleft Modelli della teoria
cinetica matematica nello studio dei sistemi complessi nelle scienze
applicate\textquotedblright \thinspace , the Consortium for Magnetofluid
Dynamics, Trieste, Italy, COST action P17 and the PDRE (Project de Recherche
Europeenne) GAMAS, CNRS, France.



\section{Appendix A: The mathematical description of incompressible NS fluids%
}

In fluid dynamics the state of an arbitrary fluid system is assumed to be
defined everywhere in a suitable \textit{extended configuration domain} $%
\Omega \times I$ [$\Omega $ \ denoting the configuration space and $%
I\subseteq
\mathbb{R}
$ the time axis] by an appropriate set of suitably smooth functions $\left\{
Z\right\} ,$ denoted as \textit{fluid fields, }and by a well-posed set of
PDE's, denoted as \textit{fluid equations, }of which the former are\textit{\
}solutions\textit{.} The fluid fields are by assumption functions of the
observables $(\mathbf{r,}t)$, with $\mathbf{r}$ and $t$ spanning
respectively the sets $\Omega $ and $I,$ namely smooth real functions.
Therefore, they are also \textit{strong solutions} of the fluid equations.
In particular, this means that they are are required to be at least
continuous in all points of the closed set $\overline{\Omega }\times I$,
with $\overline{\Omega }=\Omega \cup \partial \Omega $ closure of $\Omega ${$%
.$ \ In the remainder we shall require, for definiteness, that:}

\begin{enumerate}
\item {$\Omega ${\ (\textit{configuration domain}) is a bounded subset of
the Euclidean space }$E^{3}$ on }$%
\mathbb{R}
${$^{3};$}

\item $I$ (\textit{time axis}) is identified, when appropriate, either with
a bounded interval, \textit{i.e.}, $I${$=$}$\left] {t_{0},t_{1}}\right[
\subseteq
\mathbb{R}
,$ or with the real axis $%
\mathbb{R}
$;

\item in the open set $\Omega \times ${$I$} the functions $\left\{ Z\right\}
,$ are assumed to be solutions of a closed set of fluid equations\textit{. }%
In the case of an incompressible Navier-Stokes fluid {t}he fluid fields are $%
\left\{ Z\right\} \mathbf{\equiv }\left\{ \mathbf{V},p,S_{T}\right\} $ and
their fluid equations
\begin{eqnarray}
\rho &=&\rho _{o},  \label{1b} \\
\nabla \cdot \mathbf{V} &=&0,  \label{1ba} \\
N\mathbf{V} &=&0,  \label{1bbb} \\
\frac{\partial }{\partial t}S_{T} &=&0,  \label{1bb} \\
Z(\mathbf{r,}t_{o}) &\mathbf{=}&Z_{o}(\mathbf{r}),  \label{1ca} \\
\left. Z(\mathbf{r,}t)\right\vert _{\partial \Omega } &\mathbf{=}&\left.
Z_{w}(\mathbf{r,}t)\right\vert _{\partial \Omega },  \label{1c}
\end{eqnarray}%
where Eqs.(\ref{1b})-(\ref{1bb}) denote the{\ \textit{incompressible
Navier-Stokes equations} (INSE) and Eqs. (\ref{1b})- (\ref{1c}) the
corresponding \textit{initial-boundary value INSE problem}. In particular},
Eqs. (\ref{1b})- (\ref{1c}) are respectively the \textit{incompressibility,
isochoricity, Navier-Stokes and constant thermodynamic entropy equations}
and the initial and Dirichlet boundary conditions for $\left\{ Z\right\} ,$
with $\left\{ Z_{o}(\mathbf{r})\right\} $ and $\left\{ \left. Z_{w}(\mathbf{%
r,}t)\right\vert _{\partial \Omega }\right\} $ suitably prescribed initial
and boundary-value fluid fields, defined respectively at the initial time $%
t=t_{o}$ and on the boundary $\partial \Omega .$

\item by assumption, these equations together with appropriate initial and
boundary conditions are required to define a well-posed problem with unique
strong solution defined everywhere in $\Omega \times ${$I$}.
\end{enumerate}

Here the notation as follows. $N$ is the \textit{NS nonlinear operator}
\begin{equation}
N\mathbf{V}=\frac{D}{Dt}\mathbf{V}-\mathbf{F}_{H},  \label{NS operator}
\end{equation}%
with $\frac{D}{Dt}\mathbf{V}$ and\textbf{\ }$\mathbf{F}_{H}$ denoting
respectively the \textit{Lagrangian fluid acceleration} and the \textit{%
total force} \textit{per unit mass}
\begin{eqnarray}
&&\left. \frac{D}{Dt}\mathbf{V}=\frac{\partial }{\partial t}\mathbf{V}+%
\mathbf{V}\cdot \nabla \mathbf{V,}\right. \\
&&\left. \mathbf{F}_{H}\equiv \mathbf{-}\frac{1}{\rho _{o}}\nabla p+\frac{1}{%
\rho _{o}}\mathbf{f}+\upsilon \nabla ^{2}\mathbf{V,}\right.  \label{2c}
\end{eqnarray}%
while $\rho _{o}>0$ and $\nu >0$ are the \textit{constant mass density} and
the constant \textit{kinematic viscosity}. In particular, $\mathbf{f}$ is
the \textit{volume force density} acting on the fluid, namely which is
assumed of the form%
\begin{equation}
\mathbf{f=-\nabla }\phi (\mathbf{r},t)+\mathbf{f}_{R},
\end{equation}%
$\phi (\mathbf{r},t)$ being a suitable scalar potential, so that the first
two force terms [in Eq.(\ref{2c})] can be represented as $-\nabla p+\mathbf{f%
}$ $=-\nabla p_{r}+\mathbf{f}_{R},$ with
\begin{equation}
p_{r}(\mathbf{r},t)=p(\mathbf{r},t)-\phi (\mathbf{r},t),
\end{equation}%
denoting the \textit{reduced fluid pressure}. As a consequence of Eqs.(\ref%
{1b}),(\ref{1ba}) and (\ref{1bbb}) it follows that the fluid pressure
necessarily satisfies the \textit{Poisson equation}%
\begin{equation}
\nabla ^{2}p=S,  \label{Poisson}
\end{equation}%
where the source term $S$ reads
\begin{equation}
S=-\rho _{o}\nabla \cdot \left( \mathbf{V}\cdot \nabla \mathbf{V}\right)
+\nabla \cdot \mathbf{f}.
\end{equation}

\subsection{A.1 - Physical/conditional observables - Hidden variables}

The fluid fields $\left\{ Z\right\} $ are, by assumption, prescribed smooth
real functions of \ $(\mathbf{r},t)\in $ $\Omega \times I$. \ In particular,
they can\ be either \textit{physical observables}\textbf{\ }or \textit{%
conditional observable, }according to the definitions indicated below\textit{%
.}

\subsubsection{\textbf{Definition - P\textit{hysical observable/}conditional
observable}}

A \textit{physical observable} is an arbitrary real-valued and
uniquely-defined smooth real function of $(\mathbf{r,}t)\in $ $\Omega \times
I$. Hence, as a particular case $(\mathbf{r,}t)$ are observable too.

A \textit{conditional observable} is, instead, an arbitrary real-valued and
uniquely-defined smooth real function of $(\mathbf{r,}t)\in $ $\Omega \times
I$ which depends also on non-observable variables and is, as such, an
uniquely-prescribed function of the latter ones.

Therefore the functions $Z_{i}$ can be assumed respectively of the form \cite%
{Tessarotto2008-1,Tessarotto2009}%
\begin{equation}
Z_{i}\equiv Z_{i}(\mathbf{r},t)  \label{determistic fluid fields}
\end{equation}%
or
\begin{equation}
Z_{i}\equiv Z_{i}(\mathbf{r},t,\mathbf{\alpha }),
\label{hidden-variable fluid fields}
\end{equation}%
$\mathbf{\alpha }\in V_{\mathbf{\alpha }}\subseteq
\mathbb{R}
^{k}$ (with $k\geq 1$) denoting a suitable set of \textit{hidden variables.}
In fluid dynamics these are intended as:

\subsubsection{\textbf{Definition - Hidden variables}}

A \textit{hidden variable} is as an arbitrary real variable which is
independent of\textit{\ }$(\mathbf{r},t)$ and is not an observable$.$

\subsection{A.2 - Deterministic and stochastic fluid fields}

Hence, fluid fields of the type (\ref{hidden-variable fluid fields}) are
manifestly non-observables. However, if in the whole set $\overline{\Omega }%
\times I\times V_{\mathbf{\alpha }},$ they are uniquely-prescribed functions
of $(\mathbf{r},t,\mathbf{\alpha })$ then they are \textit{conditional
observables}. Hidden variables can be considered in principle either \textit{%
deterministic} or as \textit{stochastic variables, }in the sense specified
as follows.

\subsubsection{\textbf{Definition - Stochastic variables}}

Let $(S,\Sigma ,P)$ be a probability space; a measurable function $\mathbf{%
\alpha :}S\longrightarrow V_{\mathbf{\alpha }}$, where $V_{\mathbf{\alpha }%
}\subseteq
\mathbb{R}
^{k}$, is called \textit{stochastic} (or \textit{random}) \textit{variable}.

A stochastic variable $\mathbf{\alpha }$\ is called \textit{continuous} if%
\textit{\ }it is endowed with a \textit{stochastic model} $\left\{ g_{%
\mathbf{\alpha }},V_{\mathbf{\alpha }}\right\} ,$\textit{\ }namely a real
function\textit{\ }$g_{\mathbf{\alpha }}$ (called as \textit{stochastic PDF})%
\textit{\ }defined on the set $V_{\mathbf{\alpha }}$ and such that:

1) $g_{\mathbf{\alpha }}$ is measurable, non-negative, and of the form
\begin{equation}
g_{\mathbf{\alpha }}=g_{\mathbf{\alpha }}(\mathbf{r},t,\mathbf{\cdot });
\label{stochastic PDF}
\end{equation}

2) if $A\subseteq V_{\mathbf{\alpha }}$ is an arbitrary Borelian subset of $%
V_{\mathbf{\alpha }}$ (written $A\in \mathcal{B}(V_{\mathbf{\alpha }})$),
the integral
\begin{equation}
P_{\mathbf{\alpha }}(A)=\int\limits_{A}d\mathbf{x}g_{\mathbf{\alpha }}(%
\mathbf{r},t,\mathbf{x})  \label{dist-of-alpha}
\end{equation}%
exists and is the probability that $\mathbf{\alpha \in }A$; in particular,
since $\mathbf{\alpha }\in V_{\mathbf{\alpha }}$, $g_{\mathbf{\alpha }}$
admits the normalization
\begin{equation}
\int\limits_{V_{\mathbf{\alpha }}}d\mathbf{x}g_{\mathbf{\alpha }}(\mathbf{r}%
,t,\mathbf{x})=P_{\mathbf{\alpha }}(V_{\mathbf{\alpha }})=1.
\label{normalization}
\end{equation}

The set function $P_{\mathbf{\alpha }}:\mathcal{B}(V_{\mathbf{\alpha }%
})\rightarrow \lbrack 0,1]$ defined by (\ref{dist-of-alpha}) is a
probability measure and is called distribution (or law) of $\mathbf{\alpha }$%
. Consequently, if a function $f\mathbf{:}V_{\mathbf{\alpha }%
}\longrightarrow V_{f}\subseteq
\mathbb{R}
^{m}$ is measurable, $f$ is a stochastic variable too.

Finally define the \textit{stochastic-averaging operator }$\left\langle
\cdot \right\rangle _{\mathbf{\alpha }}$(see also \cite%
{Tessarotto2008-1,Tessarotto2009}) as\textit{\ }%
\begin{equation}
\left\langle f\right\rangle _{\mathbf{\alpha }}=\left\langle f(\mathbf{y}%
,\cdot )\right\rangle _{\mathbf{\alpha }}\equiv \int\limits_{V_{\mathbf{%
\alpha }}}d\mathbf{x}g_{\mathbf{\alpha }}(\mathbf{r},t,\mathbf{x})f(\mathbf{y%
},\mathbf{x}),  \label{stochastic averaging operator}
\end{equation}%
for any $P_{\mathbf{\alpha }}$-integrable function $f(\mathbf{y},\cdot ):V_{%
\mathbf{\alpha }}\rightarrow
\mathbb{R}
$, where the vector $\mathbf{y}$ is some parameter.

\subsubsection{\textbf{Definition - Homogeneous, stationary stochastic model}%
}

The stochastic model $\left\{ g_{\mathbf{\alpha }},V_{\mathbf{\alpha }%
}\right\} $ is denoted:

a) \textit{homogeneous }if $g_{\mathbf{\alpha }}$ is independent of $\mathbf{%
r,}$ namely
\begin{equation}
g_{\mathbf{\alpha }}=g_{\mathbf{\alpha }}(t,\mathbf{\cdot });
\label{homogeneous stoch PDF}
\end{equation}

b) \textit{stationary} if $g_{\mathbf{\alpha }}$ is independent of $t,$
i.e.,
\begin{equation}
g_{\mathbf{\alpha }}=g_{\mathbf{\alpha }}(\mathbf{r},\mathbf{\cdot }).
\label{stationary stoch PDF}
\end{equation}

\subsubsection{\textbf{Definition - Deterministic variables}}

Instead, if $g_{\mathbf{\alpha }}(\mathbf{r},t,\mathbf{\cdot })$ is a
\textit{deterministic PDF}, namely it is of the form%
\begin{equation}
g_{\mathbf{\alpha }}(\mathbf{r},t,\mathbf{x})=\delta ^{(k)}(\mathbf{x-\alpha
}_{o}),  \label{deterministic g}
\end{equation}

$\delta ^{(k)}(\mathbf{x-\alpha }_{o})$ denoting the $k$-dimensional Dirac
delta in the space $V_{\mathbf{\alpha }},$ the hidden variables $\mathbf{%
\alpha }$ are denoted as \textit{deterministic}.

Let us now assume that, for a suitable stochastic model $\left\{ g_{\mathbf{%
\alpha }},V_{\mathbf{\alpha }}\right\} $, with $g_{\mathbf{\alpha }}$
non-deterministic, the stochastic variables $Z_{i}\equiv Z_{i}(\mathbf{r},t,%
\mathbf{\alpha })$ a{nd } $f_{1}(\mathbf{r,v},t,\mathbf{\alpha })$ (where $%
Z_{i}(\mathbf{r},t,\mathbf{\cdot })$ a{nd } $f_{1}(\mathbf{r,v},t,\mathbf{%
\cdot })${\ are measurable functions) a}dmit everywhere in $\overline{\Omega
}\times I$ and $\overline{\Gamma }\times I$ $\ $the \textit{stochastic
averages }$\left\langle Z_{i}\right\rangle _{\mathbf{\alpha }}$ and $%
\left\langle f_{1}\right\rangle _{\mathbf{\alpha }}$ defined by (\ref%
{stochastic averaging operator}).

Hence, $Z_{i}\equiv Z_{i}(\mathbf{r},t,\mathbf{\alpha }),$ $f_{1}(\mathbf{r,v%
},t,\mathbf{\alpha })$ and the mean-field force $\mathbf{F}(f_{1})$\textbf{\
}[see Sections 2,3 and 4]\textbf{\ }admit also the \textit{stochastic
decompositions}%
\begin{eqnarray}
Z_{i} &=&\left\langle Z_{i}\right\rangle _{\mathbf{\alpha }}+\delta Z_{i},
\label{stochastic decomposition} \\
f_{1} &=&\left\langle f_{1}\right\rangle _{\mathbf{\alpha }}+\delta f_{1},
\label{stochastic decomposition 2} \\
\mathbf{F}(f_{1}) &=&\left\langle \mathbf{F}(f_{1})\right\rangle _{\mathbf{%
\alpha }}+\delta \mathbf{F}(f_{1}).  \label{stochastic decomposition 3}
\end{eqnarray}

In particular, unless $g_{\mathbf{\alpha }}(\mathbf{r},t,\mathbf{\cdot })$
is suitably smooth, it follows that generally $\left\langle
Z_{i}\right\rangle _{\mathbf{\alpha }},\delta Z_{i}$ and respectively $%
\left\langle f_{1}\right\rangle _{\mathbf{\alpha }},\delta f_{1}$ may belong
to different functional classes with respect to the variables $(\mathbf{r}%
,t) $.

\subsection{A.3 - Deterministic and stochastic INSE problems - Regular and
turbulent flows}

Therefore, assuming, for definiteness, that all the fluid fields $Z,$ the
volume force $\mathbf{f}$ and the initial and boundary conditions, are
either deterministic or stochastic variables and both belong to the same
functional class, i.e., are suitably smooth w.r. to $(\mathbf{r},t)$ and $%
\mathbf{\alpha }$,\ Eqs. (\ref{1b})- \ref{1c}) define respectively a \textit{%
deterministic} or \textit{stochastic initial-boundary value INSE problem}.
In both cases we shall assume that it admits a strong solution in $\overline{%
\Omega }\times I$ (or $\overline{\Omega }\times I\times V_{\alpha }$).

In the first case, which characterizes flows to be denoted as \textit{regular%
}, the fluid fields are by assumption \textit{physical observables,} i.e.,
uniquely-defined, smooth, real functions of $\ (\mathbf{r,}t)\in $ $\Omega
\times I$ \ [with $\Omega ,$ the \textit{configuration space,} and\textit{\ }%
$\overline{\Omega }$ its closure, to be assumed subsets of the Euclidean
space on\textit{\ }$%
\mathbb{R}
^{3}$ and $I,$ the \textit{time axis,} denoting a subset of $%
\mathbb{R}
$].

In the second case, characterizing instead \textit{turbulent flows}, the
fluid fields are only \textit{conditional observables} (see again Subsection
A.1). In this case, besides $\left( \mathbf{r},t\right) $, they may be
assumed to depend also on a suitable stochastic variable $\mathbf{\alpha }$,
(with $\mathbf{\alpha \in }V_{\mathbf{\alpha }}$ and $V_{\mathbf{\alpha }}$
subset of $%
\mathbb{R}
^{k}$ with $k\geq 1$).\ Hence they are stochastic variables too.

\section{Appendix B: definition of $N_{1}$}

Let us define provide here an explicit definition of $N_{1}(\mathbf{r}_{i}%
\mathbf{,v,}t,\mathbf{\alpha };Z)$ [required to specify also $\widehat{f}%
_{1}^{(freq)}$ in terms of Eq.(\ref{stat-1})]. For definiteness, let us
assume that the fluid velocity is bounded, i.e., that there exists $V_{B}\in
\mathbb{R}
$ such that in $\Omega \times I$ for each component of the fluid velocity $%
V_{k}(\mathbf{r,}t,\mathbf{\alpha })$ (with $k=1,2,3$) there results
\begin{equation}
\left\vert V_{k}(\mathbf{r,}t,\mathbf{\alpha })\right\vert \leq \frac{1}{2}%
V_{B}.  \label{B.1}
\end{equation}%
Then $N_{1}(\mathbf{r}_{i}\mathbf{,v,}t,\mathbf{\alpha };Z)$ can be defined
as follows%
\begin{eqnarray}
&&N_{1}(\mathbf{r}_{i}\mathbf{,v,}t,\mathbf{\alpha };Z)=  \label{B.3} \\
&=&\frac{N}{c}\prod\limits_{k=1,2,3}\Theta _{ik}(\mathbf{v})\Theta (\frac{%
V_{B}^{2}}{4}\left[ 1-\frac{1}{M}\right] ^{2}-v_{k}^{2}), \\
&&\Theta _{ik}(\mathbf{v})\equiv \Theta (V_{k}(\mathbf{r}_{i},t)-v_{k}-\frac{%
V_{B}}{2M})  \label{B.4} \\
&&\Theta (v_{k}-V_{k}(\mathbf{r}_{i},t)+\frac{V_{B}}{2M}),  \notag
\end{eqnarray}%
with $M\in
\mathbb{N}
$ and $\Theta (x)$ the Heaviside theta function$;$ here $c\in
\mathbb{R}
$ and $N\in
\mathbb{N}
$ are defined so that there results%
\begin{eqnarray}
c &=&V_{B}^{3}\sum\limits_{i=1,N}\prod\limits_{k=1,2,3}\Theta _{ik}(\mathbf{v%
}),  \label{B.5} \\
M^{3} &=&N.  \label{B.6}
\end{eqnarray}%
Thanks to positions (\ref{B.3})-(\ref{B.6}) for an arbitrary $N\in
\mathbb{N}
$ fulfilling Eq.(\ref{B.6}), it follows
\begin{equation}
\lim_{N\rightarrow \infty }\frac{1}{N}\sum\limits_{i=1,N}\int%
\limits_{U}d^{3}vN_{1}(\mathbf{r}_{i}\mathbf{,v,}t,\mathbf{\alpha };Z)=1.
\label{B.7}
\end{equation}%
Hence Eq.(\ref{stat-2}) is satisfied identically.

\section{Appendix C: evaluation of $\mathbf{F}^{(T)}$}

In $f_{1}(1)$ and $f_{1}(2)$ coincide with a local Gaussian [i.e., see Eq.(%
\ref{Maxwellian case})] there results by construction%
\begin{eqnarray}
\int d^{3}\mathbf{VV}f_{1}(1)f_{1}(2) &=&\left[ \beta \mathbf{v-U}\right]
f_{2}(\mathbf{r}_{1},\mathbf{r}_{2},\mathbf{v;}Z) \\
\int d^{3}\mathbf{V}V^{2}f_{1}(1)f_{1}(2) &=&\frac{3}{2}v_{th}^{2}f_{2}(%
\mathbf{r}_{1},\mathbf{r}_{2},\mathbf{v;}Z),  \label{int-2}
\end{eqnarray}%
where there results%
\begin{eqnarray}
\beta &=&\frac{v_{th,p}^{2}(1)-v_{th,p}^{2}(2)}{%
v_{th,p}^{2}(1)+v_{th,p}^{2}(2)}=  \label{int-3} \\
&&\left. =\frac{1}{\frac{1}{v_{th,p}^{2}(2)}-\frac{1}{v_{th,p}^{2}(1)}}%
,\right.  \notag \\
\mathbf{U} &\mathbf{=}&\frac{\mathbf{V}(1)v_{th,p}^{2}(2)+\mathbf{V}%
(2)v_{th,p}^{2}(1)}{v_{th,p}^{2}(1)+v_{th,p}^{2}(2)}=  \label{int-4} \\
&=&\frac{\frac{\mathbf{V}(1)}{v_{th,p}^{2}(1)}+\frac{\mathbf{V}(2)}{%
v_{th,p}^{2}(2)}}{\frac{1}{v_{th,p}^{2}(2)}-\frac{1}{v_{th,p}^{2}(1)}}.
\notag
\end{eqnarray}%
Let us now evaluate the expression $\int d^{3}\mathbf{V}\frac{1}{2}\left[
\mathbf{F}(1)-\mathbf{F}(2)\right] f_{1}(1)f_{1}(2).$ Introducing the
notations%
\begin{eqnarray}
\mathbf{F}_{0}(1) &=&\frac{1}{\rho _{0}}\mathbf{f}(1)+\left[ \mathbf{V+v-V}%
(1)\right] \cdot \nabla _{1}\mathbf{V}\mathbb{(}1\mathbb{)}+ \\
&&+\nu \nabla _{1}^{2}\mathbf{V}(1)\mathbf{,}  \notag
\end{eqnarray}%
\begin{eqnarray}
&&\left. \mathbf{F}_{1}\mathbf{(}1)=\frac{\left[ \mathbf{V+v-V}(1)\right] }{2%
}A(1)+\right. \\
&&+\frac{v_{th,p}^{2}(1)}{2p_{1}(1)}\mathbf{\nabla }_{1}p_{1}(1)\left\{
\frac{\left[ \mathbf{V+v-V}(1)\right] ^{2}}{v_{th,p}^{2}(1)}-\frac{3}{2}%
\right\} ,  \notag
\end{eqnarray}%
\begin{eqnarray}
\mathbf{F}_{0}(2) &=&\frac{1}{\rho _{0}}\mathbf{f}(2)+\left[ \mathbf{V-v-V}%
(2)\right] \cdot \nabla _{2}\mathbf{V(2)+} \\
&&+\nu \nabla _{2}^{2}\mathbf{V(2),}  \notag
\end{eqnarray}%
\begin{eqnarray}
&&\left. \mathbf{F}_{1}\mathbf{(}2)=\frac{\left[ \mathbf{V-v-V}(1)\right] }{2%
}A(2)+\right. \\
&&+\frac{v_{th,p}^{2}(2)}{2p_{1}(2)}\mathbf{\nabla }_{2}p_{1}(2)\left\{
\frac{\left[ \mathbf{V-v-V}(2)\right] ^{2}}{v_{th,p}^{2}(2)}-\frac{3}{2}%
\right\} ,  \notag
\end{eqnarray}%
and letting for $j=1,2,$
\begin{equation}
A\mathbf{(}j)=\frac{1}{p_{1}(j)}\frac{D}{Dt}p_{1}(j)
\end{equation}%
it follows%
\begin{eqnarray*}
&&\left. \int d^{3}\mathbf{V}\frac{1}{2}\left[ \mathbf{F}(1)-\mathbf{F}(2)%
\right] f_{1}(1)f_{1}(2)=\right. \\
&=&\left[ \mathbf{F}_{0}^{(T)}+\mathbf{F}_{1}^{(T)}\right] f_{2}\equiv
\mathbf{F}^{(T)}f_{2},
\end{eqnarray*}%
where%
\begin{eqnarray}
&&\left. 2\mathbf{F}_{0}^{(T)}=\frac{1}{\rho _{0}}\mathbf{f}(1)+\right.
\label{F0-B-} \\
&&+\left[ \left( \beta +1\right) \mathbf{v-U-V}(1)\right] \cdot \nabla _{1}%
\mathbf{V}\mathbb{(}1\mathbb{)}\mathbf{+}  \notag \\
&&+\nu \nabla _{1}^{2}\mathbf{V}(1)-  \notag \\
&&-\frac{1}{\rho _{0}}\mathbf{f}(2)-  \notag \\
&&-\left[ \left( \beta +1\right) \mathbf{v-U-V}(2)\right] \cdot \nabla _{2}%
\mathbf{V}\mathbb{(}2\mathbb{)}\mathbf{+}\nu \nabla _{2}^{2}\mathbf{V}(2)
\notag
\end{eqnarray}%
\begin{eqnarray}
&&\left. 2\mathbf{F}_{1}^{(T)}=\frac{\left[ \left( \beta +1\right) \mathbf{%
v-V}(1)\right] }{2}A(1)+\right.  \label{F1-B-} \\
&&+\frac{v_{th,p}^{2}(1)}{2p_{1}(1)}\mathbf{\nabla }_{1}p_{1}(1)  \notag \\
&&\left\{ \frac{\frac{3}{2}v_{th}^{2}+2\left[ \beta \mathbf{v-U}\right]
\cdot \left[ \mathbf{v-V}(1)\right] +\left[ \mathbf{v-V}(1)\right] ^{2}}{%
v_{th,p}^{2}(1)}-\frac{3}{2}\right\} -  \notag \\
&&-\frac{\left[ \left( \beta +1\right) \mathbf{v-V}(2)\right] }{2}A(2)-
\notag \\
&&-\frac{v_{th,p}^{2}(2)}{2p_{1}(2)}\mathbf{\nabla }_{2}p_{1}(2)  \notag \\
&&\left\{ \frac{\frac{3}{2}v_{th}^{2}+2\left[ \beta \mathbf{v-U}\right]
\cdot \left[ \mathbf{v-V}(2)\right] +\left[ \mathbf{v-V}(2)\right] ^{2}}{%
v_{th,p}^{2}(2)}-\frac{3}{2}\right\} .  \notag
\end{eqnarray}


\end{document}